\documentclass[reprint, aps, prd, letterpaper, noshowpacs, amsmath, %
amssymb, amsfonts, nofootinbib, floatfix, 
twoside]{revtex4-1}
\pdfoutput=1

\usepackage[T3,T1]{fontenc}
\usepackage{mathtools}
\usepackage{bm} 

\usepackage{graphicx} %
\usepackage{color} %
\usepackage{xspace}
\usepackage{braket}
\usepackage{accents}
\usepackage{siunitx}

\definecolor{CiteColor}{rgb}{0.18039, 0.18824, 0.57255}
\definecolor{UrlColor} {rgb}{0.741, 0.173, 0.000}
\definecolor{LinkColor}{rgb}{0.25098, 0.47843, 0.04706}

\usepackage[normalem]{ulem}

\makeatletter

\usepackage[colorlinks, plainpages=false,
hyperfigures=true]{hyperref}
\hypersetup{linkcolor=LinkColor}
\hypersetup{citecolor=CiteColor}
\hypersetup{urlcolor=UrlColor}

\makeatother

\newcommand{\etal}{\textit{et~al.}\@\xspace}
\newcommand{\etalp}{\textit{et~al.}\xspace}
\newcommand{\pN}{pN\xspace}

\newcommand{\LN}{\ensuremath{\hat{L}_{\text{N}}}}
\newcommand{\LNdot}{\ensuremath{\dot{\hat{L}}_{\text{N}}}}

\let\Originalddefinition\d
\renewcommand{\d}{\ensuremath{\mathrm{d}}}
\newcommand{\e}{\ensuremath{\mathrm{e}}}
\let\Originalidefinition\i
\renewcommand{\i}{\ensuremath{\mathrm{i}}}

\newcommand{\tfid}{\ensuremath{t_\text{fid}}}
\DeclareMathOperator{\arccot}{arccot}

\let\Originalcdefinition\c
\renewcommand{\c}{\mathrm{c}}

\newcommand{\MSun}{\ensuremath{M_\odot}\xspace}

\newcommand{\abs} [1]{\left\lvert{#1}\right\rvert}

\newcommand{\co}[1]{\ensuremath{\bar{#1}}}

\newcommand{\h}{\ensuremath{h} }

\newcommand{\sYlm}[1]{\ensuremath{\scripts{_{s}}{Y}{_{#1}}}}

\DeclareSymbolFont{tipa}{T3}{tipa}{m}{n}
\DeclareMathAccent{\ibreve}{\mathalpha}{tipa}{'020}
\newcommand{\Rotated}[1]{\ensuremath{\ibreve{#1}}}
\newcommand{\Rotation}[1]{\ensuremath{\mathbf{#1}}}
\newcommand{\Generator}[1]{\ensuremath{\MakeUppercase{#1}}}

\newcommand{\SO}[1]{\ensuremath{SO(#1)}}
\newcommand{\so}[1]{\ensuremath{\mathfrak{so}(#1)}}

\newcommand{\wavefield}{q}
\newcommand{\capwavefield}{Q}
\newcommand{\FrameRot}{\ensuremath{\vec{\varpi}}}
\newcommand{\FrameRotGen}{\ensuremath{\Generator{\varPi}}}
\newcommand{\YawfulRot}{\ensuremath{\Rotation{R}_{\text{ra}}}}
\newcommand{\YawfulRotDot}{\ensuremath{\dot{\Rotation{R}}_{\text{ra}}}}
\newcommand{\YawfulRotQ}{\ensuremath{R_{\text{ra}}}}
\newcommand{\YawfulRotDotQ}{\ensuremath{\dot{R}_{\text{ra}}}}
\newcommand{\YawfulRotCoQ}{\ensuremath{\co{R}_{\text{ra}}}}
\newcommand{\GammaRot}{\ensuremath{\Rotation{R}_{\gamma}}}
\newcommand{\GammaRotQ}{\ensuremath{R_{\gamma}}}

\newcommand{\RFrame}{\ensuremath{\Rotation{R}_{\text{f}}}}

\makeatletter

\newcommand{\ShowDimensions}{%
  \typeout{The font encoding is \f@encoding}        %
  \typeout{The font family is \f@family}            %
  \typeout{The font series is \f@series}            %
  \typeout{The font shape is \f@shape}              %
  \typeout{The font size is \f@size}                %
  \typeout{The baselineskip is \f@baselineskip}     %
  \typeout{The math font size is \tf@size}          %
  \typeout{The math script size is \sf@size}        %
  \typeout{The math scriptscript size is \ssf@size} %
  \typeout{The linewidth is \the\linewidth}         %
}
\newcommand{\prefixscripts}[2]{%
  \@mathmeasure\z@\displaystyle{#2}%
  \global\setbox\@ne\vbox to\ht\z@{}\dp\@ne\dp\z@
  \setbox\tw@\box\@ne
  \@mathmeasure4\displaystyle{\copy\tw@#1}%
  \@mathmeasure6\displaystyle{#2}%
  \dimen@-\wd6 \advance\dimen@\wd4 \advance\dimen@\wd\z@
  \hbox to\dimen@{}{\kern-\dimen@\box4\box6}%
}
\newcommand{\scripts}[3]{%
  \@mathmeasure\z@\displaystyle{#2}%
  \global\setbox\@ne\vbox to\ht\z@{}\dp\@ne\dp\z@
  \setbox\tw@\box\@ne
  \@mathmeasure4\displaystyle{\copy\tw@#1}%
  \@mathmeasure6\displaystyle{#2#3}%
  \dimen@-\wd6 \advance\dimen@\wd4 \advance\dimen@\wd\z@
  \hbox to\dimen@{}{\kern-\dimen@\box4\box6}%
}
\makeatother

\makeatletter
\let\protect\relax
{\catcode`\|=\active
  \xdef\InnerProduct{\protect\expandafter\noexpand\csname InnerProduct \endcsname}
  \expandafter\gdef\csname InnerProduct \endcsname#1{%
    \begingroup
    \ifx\SavedDoubleVert\relax
    \let\SavedDoubleVert\|\let\|\IpDoubleVert
    \fi
    \mathcode`\|32768\let|\IPVert
    \left({#1}\right)
    \endgroup
  }
}
\def\IPVert{\@ifnextchar|{\|\@gobble}
     {\egroup\,\mid@vertical\,\bgroup}}
\def\IPDoubleVert{\egroup\,\mid@dblvertical\,\bgroup}
\let\SavedDoubleVert\relax
\def\midvert{\egroup\mid\bgroup}
\def\SetVert{\@ifnextchar|{\|\@gobble}
    {\egroup\;\mid@vertical\;\bgroup}}
\def\SetDoubleVert{\egroup\;\mid@dblvertical\;\bgroup}
\def\mid@vertical{\mskip1mu\vrule\mskip1mu}
\def\mid@dblvertical{\mskip1mu\vrule\mskip2.5mu\vrule\mskip1mu}
\makeatother

\usepackage[multidot]{grffile} 

\newcommand{\Cornell}{\affiliation{Center for Radiophysics and Space
    Research, Cornell University, Ithaca, New York 14853, USA}}
\newcommand{\CITA}{\affiliation{Canadian Institute for Theoretical 
    Astrophysics, University of Toronto, Toronto, Ontario M5S 3H8,
    Canada}}

\begin{document}

\title{A geometric approach to the precession of compact binaries}

\author{Michael Boyle} %
\author{Robert Owen} %
\Cornell %
\author{Harald P. Pfeiffer} %
\CITA %

\date{\today}

\begin{abstract}
  We discuss a geometrical method to define a preferred reference
  frame for precessing binary systems and the gravitational waves they
  emit.  This minimal-rotation frame is aligned with the
  angular-momentum axis and fixes the rotation about that axis up to a
  constant angle, resulting in an essentially invariant frame.
  Gravitational waveforms decomposed in this frame are similarly
  invariant under rotations of the inertial frame and exhibit
  relatively smoothly varying phase.  By contrast, earlier
  prescriptions for radiation-aligned frames induce extraneous
  features in the gravitational-wave phase which depend on the
  orientation of the inertial frame, leading to fluctuations in the
  frequency that may compound to many gravitational-wave cycles.  We
  explore a simplified description of post-Newtonian approximations
  for precessing systems using the minimal-rotation frame, and
  describe the construction of analytical/numerical hybrid waveforms
  for such systems.
\end{abstract}

\pacs{%
  04.30.-w, 
  04.80.Nn, 
  04.25.D-, 
  04.25.dg 
}


\maketitle

\section{Introduction}
\label{sec:Introduction}

One of the central goals of modern numerical relativity is the
accurate simulation of compact binary systems, in particular the
computation of the gravitational waveforms emitted by these systems.
These waveforms provide crucial input into the construction of
accurate template banks necessary for detection and parameter
estimation based on matched filtering~\cite{Finn:1992,
  FinnChernoff:1993} in gravitational-wave detectors such as LIGO,
Virgo, LCGT~\cite{Barish:1999, Sigg:2008, Acernese:2008, Kuroda:2010},
and possible space-based detectors such as LISA~\cite{Lisa, Lisa98,
  Jennrich:2009}.  More generally, detailed and accurate knowledge of
waveforms provides a dictionary to relate measured waveforms to the
specific details of the astrophysical sources that give rise to those
waveforms, allowing gravitational-wave experiments to fulfill their
proper role as tools for extremely high-precision astrophysics.

For such a bank of gravitational waveforms to be useful, however, it
must not be restricted to an astrophysically unrealistic subset of the
space of source parameters.  Numerical-relativity simulations must
eventually treat binary systems with a broad range of mass ratios,
spin magnitudes, and spin orientations.  Many of the fundamental
challenges on the first two points have now been
overcome.\footnote{See reviews~\cite{Centrella:2010,
    McWilliams:2010iq}, as well as \cite{LoustoZlochower2010,
    Lovelace2010, Lovelace:2011} for simulations that push the limits
  of large mass ratios and spins.}  However, precessing binaries
remain a formidable challenge.  The parameter space of precessing
binaries is vastly larger than for non-precessing binaries, and its
exploration is just getting underway.  Furthermore, the absence of
precession allows simplifying assumptions about the properties of the
gravitational waveforms, greatly easing post-Newtonian comparisons and
simplifying gravitational-wave data-analysis strategies.

When binary systems of black holes (or neutron stars or other compact
objects) have spin angular momenta that are misaligned (with one
another or with the orbital angular momentum of the pair), the plane
of orbital motion inclines and precesses over time.  In post-Newtonian
theory, this phenomenon can be interpreted as arising from spin-orbit
couplings.  In fully nonlinear general relativity, these dynamical
effects cannot easily be understood in gauge-unambiguous language.
However, the effects of this precession can be seen unambiguously in
the waveform, as modulations that trade energy content between the
various spherical-harmonic modes~\cite{ApostolatosEtAl:1994,
  Kidder:1995, GualtieriEtAl:2008}.

These modulations present difficulties for cataloging the
gravitational waveforms.  In non-precessing simulations, the standard
practice has been to decompose each spherical harmonic component of
the waveform into a time-varying amplitude and phase.  Both of these
elements, in non-precessing cases, can be accurately approximated
before or after merger as simple polynomials or exponentially damped
polynomials, greatly simplifying their description.  In precessing
systems, this is no longer the case.

Aside from the complexity of fitting the precessing waveforms, there
is also the concern that a precessing system does not have a preferred
frame of inertial coordinates.  An overall rotation of the inertial
coordinates would transform the various waveform modes, modulating
them in different ways.  This means that comparisons between waveforms
must account for a rotation between the inertial frames in which they
are measured.  In non-precessing cases, this difficulty is avoided by
the existence of a preferred, fixed axis (the axis along which the
radiation is preferentially beamed).  This axis is intuitively
associated with the normal to the orbital plane.

Schmidt \etal~\cite{SchmidtEtAl:2011} pointed out that this preferred
axis, while no longer fixed in precessing cases, nonetheless still
exists and can be used to rotate the inertial spatial coordinates over
time, demodulating the waveform.  They suggested a method for finding
that axis, as did O'Shaughnessy \etal~\cite{OShaughnessyEtAl:2011}
more recently.  Essentially, the idea is to find a rotation operator
at each instant in time to maximize the $z$ component of the angular
momentum in the radiation.\footnote{More specifically, the method of
  Schmidt \etal calls for a rotation that maximizes the power in the
  $(\ell, m) = (2, \pm 2)$ modes.  However, we show in
  Sec.~\ref{s:LocatingRadiationAxis} that this can be regarded---in
  some sense---as a restriction of the method of O'Shaughnessy \etalp}
The $z$ axis of the original frame, rotated by this operator, is taken
to be the preferred axis.  We refer to this axis as the
\textit{radiation axis}.

We expect this insight to be important for understanding and
cataloging generic gravitational waveforms.  Definition of the
radiation axis is the first step toward frames adapted to precessing
binaries, in which the gravitational waveforms have simple structure.
However, fixing the \emph{axis} does not fix the \emph{frame}, because
of the ambiguity in rotations about the axis.  From a more formal
perspective, the (directed) axis being tracked lives in a
two-dimensional space: the space of unit vectors in $\mathbb{R}^3$,
which is topologically the two sphere $S^2$.  However the space of
available rotations, the group manifold of $\SO3$, is topologically
$\mathbb{RP}^3$, which is three dimensional.  In general, no
mathematically preferred method exists to infer a unique path in
$\mathbb{RP}^3$ from a path in $S^2$; additional conditions must be
imposed.

Therefore, the second step toward adapted frames is to fix this
rotation about the radiation axis.  References~\cite{SchmidtEtAl:2011}
and~\cite{OShaughnessyEtAl:2011} address the first step by providing
suitable definitions of the radiation axis, but deal with the second
step only implicitly through the choice of parameterization of the
rotation matrices.  A rotation around the radiation axis changes the
phase of each gravitational-radiation mode by an integer multiple of
the rotation angle.  An unsuitable choice of this angle will induce
unphysical variations in the gravitational-wave phases in the adapted
frame, even for vanishingly small precession, as demonstrated in
Sec.~\ref{sec:EffectsOfMinimalRotation}.

The present paper addresses the question of how to fix the rotation
about the radiation axis.  Our construction is geometric, and the
resulting \textit{minimal-rotation frame} is invariant under rotations
of the inertial coordinates in which the precessing waveforms are
extracted, except for the remaining freedom of a constant overall
rotation.  Therefore, the approach proposed here results in an
essentially unique adapted frame and in gravitational waveforms that
are similarly unique.  In contrast, the implementations of
Refs.~\cite{SchmidtEtAl:2011} and~\cite{OShaughnessyEtAl:2011} choose
the final rotation angle in a coordinate-dependent manner: working in
terms of Euler angles and always setting the third Euler angle to
$\gamma=0$---by construction, this is the rotation about the $z$ axis
of the adapted frame.

The key to fixing this remaining freedom by geometric means lies in an
analogy with the non-precessing binary.  In the non-precessing case,
there are again many coordinate frames that preserve the condition
that the radiation is primarily quadrupolar with $m = \pm 2$.  One
could arbitrarily rotate about the $z$ axis and preserve this
condition.  In practice this does not pose a problem, because it is
taken as physically obvious that one would not analyze the waveform in
a coordinate frame that is rotating about the $z$ axis relative to an
inertial frame.  A \emph{fixed} overall rotation about the $z$ axis is
allowed, leading to the well-known overall freedom in the waveform
phase.  However a time-dependent rotation about the $z$ axis, which
could cause arbitrary frequency modulation in the waveform, is
rejected as unnatural.

Any frame that tracks the radiation axis of a precessing binary, on
the other hand, is necessarily changing.  As far as possible, we would
like to carry over the non-rotating condition from the non-precessing
system.  In this case, we can describe the rotation of the frame by
the instantaneous rotation vector $\FrameRot$.  Relative to an
inertial frame, the time derivative of any vector stationary in the
rotating frame is given by
\begin{equation}
  \label{eq:StationaryToRotatingVectorDeriv}
  \dot{\vec{v}} = \FrameRot \times \vec{v}~.
\end{equation}
If we denote the radiation axis by a unit vector $\vec{a}$, we see
that $\dot{\vec{a}} = \FrameRot \times \vec{a}$.  Taking the cross
product of both sides of this equation with $\vec{a}$ and using the
standard triple-product formula, we have
\begin{equation}
  \label{eq:ACrossAdot}
  \vec{a} \times \dot{\vec{a}} = (\vec{a} \cdot \vec{a})\, \FrameRot -
  (\vec{a} \cdot \FrameRot)\, \vec{a}~.
\end{equation}
Using the fact that $\vec{a}$ is unit, we can rearrange this as
\begin{equation}
  \label{eq:FrameRotationVector}
  \FrameRot = \vec{a} \times \dot{\vec{a}} + (\vec{a} \cdot
  \FrameRot)\, \vec{a}~.
\end{equation}
Of course, the component of $\FrameRot$ along $\vec{a}$ is completely
undetermined by this equation; we need some other condition to fix it.
Now, when $\dot{\vec{a}} = 0$, as in the non-precessing case, we
recover the natural non-rotating frame when $\vec{a} \cdot \FrameRot =
0$.  This is the same condition imposed by Buonanno, Chen, and
Vallisneri~\cite{BuonannoEtAl:2003} in the context of post-Newtonian
template waveforms (discussed further in
Sec.~\ref{sec:PostNewtonianWaveforms} below).  We stress the
importance of this condition more broadly---and particularly in the
context of numerical relativity.

Here, we augment the methods of Schmidt \etal and O'Shaughnessy
\etal with the condition that the instantaneous rotation of the frame
satisfy
\begin{equation}
  \label{eq:MinimalRotationCondition}
  \FrameRot \cdot \vec{a} = 0~.
\end{equation}
Hereafter, we refer to this as the condition of \textit{minimal
  rotation}, as this implies that $\FrameRot$ has the smallest
possible magnitude, out of the infinitely many rotation vectors
consistent with the known motion of the radiation axis.  It is
significant that this condition on $\FrameRot$ is \emph{geometrically}
meaningful, because $\vec{a}$ is---at any instant---independent of the
orientation of the frame in which it is found.  As we will demonstrate
below, the waveform decomposed in such a frame is independent of an
overall rotation, up to a constant phase.

Given a rotation $\Rotation{R}(t)$ that takes the $z$ axis into the
radiation axis ($\vec{a}(t)= \Rotation{R}(t)\, \hat{z}$), we can use
Eq.~\eqref{eq:MinimalRotationCondition} to find a condition on
$\Rotation{R}(t)$ that holds only if it is a \emph{minimal} rotation.
Alternatively, given any rotation that takes the $z$ axis into the
radiation axis, we can easily construct another rotation that does the
same while also satisfying the minimal-rotation condition.  These
relations are the key results of this paper.

The remainder of this paper is structured as follows: In
Sec.~\ref{s:LocatingRadiationAxis} we summarize the algorithms for
finding the radiation axis presented by Schmidt
\etal~\cite{SchmidtEtAl:2011} and by O'Shaughnessy
\etal~\cite{OShaughnessyEtAl:2011}.  We show that the first method is
essentially a restriction of the second, but point out that with
slight improvements to the numerical techniques both can be used find
the correct radiation axis to very high accuracy---at least for simple
toy models in which the correct axis is known.  In
Sec.~\ref{s:FindingMinimalRotFrame} we translate the minimal-rotation
condition, Eq.~\eqref{eq:MinimalRotationCondition}, into a condition
on the rotation operator itself, and construct a method for imposing
this condition while leaving the radiation axis fixed.  In
Sec.~\ref{sec:EffectsOfMinimalRotation}, we compare the original
algorithms to this coordinate-independent method.  First, we show that
the motion of the coordinate axes is essentially invariant for our
implementation, while the axis motion for the original methods depends
sensitively on the orientation of the inertial coordinate frame.  We
then demonstrate that this dependence shows up in phase the waveform
modes using a simple post-Newtonian model.  In
Sec.~\ref{sec:Applications}, we exhibit applications of the
minimal-rotation frame, which demonstrate that the usual machinery
used for non-precessing waveforms can be directly carried over to
precessing systems in this frame.  In particular, we discuss a
framework for calculating post-Newtonian waveforms taking advantage of
this simple frame, first proposed by Buonanno, Chen, and
Vallisneri~\cite{BuonannoEtAl:2003}.  We then describe how to compare
waveforms and construct hybrids.  Finally, in Sec.~\ref{s:Discussion},
we close with discussion of the benefits of this method and potential
applications in analytic constructions.  Two appendices detail our
conventions, list some crucial formulas for rotations, and repeat our
main results in the language of quaternions.

\section{Locating the radiation axis}
\label{s:LocatingRadiationAxis}

The gravitational waves radiated from a compact binary are typically
decomposed in a spin-weighted spherical harmonic expansion of the
field on a sphere.  For a binary with orbital angular momentum
$\vec{L}$ along the $z$ axis, the dominant modes in this expansion are
the $(\ell, m) = (2, \pm 2)$ modes.  When $\vec{L}$ is not along the
$z$ axis, however, the various modes will mix, and other modes of the
$\ell=2$ component can dominate.  For precessing binaries, this
misalignment of the angular momentum and the $z$ axis of an inertial
frame is inevitable, complicating comparisons between simulations
produced with even slightly different initial conditions.  Moreover,
the amplitude and phase of the modes themselves will become rapidly
varying functions of time, complicating analysis of the waveforms.
Both of these complications can be eliminated by decomposing the
waveform in a non-inertial frame that somehow tracks the motion of the
binary.  Two methods to do this have been presented in the literature.
We now review these, showing that they can be expressed in very
similar ways, noting that both can be implemented numerically to
achieve very high accuracy, and highlighting the crucial degeneracy
present in both.

\subsection{The two methods}
\label{sec:TheTwoMethods}
The first algorithm, presented in Schmidt \textit{et
  al.}~\cite{SchmidtEtAl:2011}, finds a frame in which the amplitudes
of the $(\ell, m) = (2, \pm 2)$ modes of the gravitational-wave field
are maximized.  Reference~\cite{SchmidtEtAl:2011} uses the
Newman-Penrose Weyl scalar $\Psi_4$, though the principle is the same
for any radiation field---for example, the metric perturbation $\h$.
We will use the generic symbol $\wavefield$ to establish the method.
We regard $\wavefield$ as the quantity measured in an inertial frame,
and $\Rotated{\wavefield}$ the function under a specified
rotation.\footnote{To be precise, we define $\Rotated{\wavefield}$ to
  be the function satisfying $\Rotated{\wavefield}(\vartheta',
  \varphi') = \wavefield(\vartheta, \varphi)$, for any angles related
  by $ \Rotation{R}(0, \vartheta', \varphi') = \Rotation{R}(\alpha,
  \beta, \gamma)\, \Rotation{R}(0, \vartheta, \varphi)$, where each
  $\Rotation{R}$ is a rotation operator parameterized by the Euler
  angles as described in Appendix~\ref{sec:Conventions}.  These
  conventions affect details of later results---for example, the form
  of Eq.~\eqref{eq:MaximizedQuantity} and its independence of $\gamma$
  (as opposed to $\alpha$).}  Each of these can be decomposed in
spin-weighted spherical harmonics [Eq.~\eqref{eq:SWSHDecomposition}],
with weights of the modes denoted $\wavefield^{\ell, m}$ and
$\Rotated{\wavefield}^{\ell, m}$.  The relation between
$\wavefield^{\ell, m}$ and $\Rotated{\wavefield}^{\ell,m}$ is given by
Eq.~\eqref{eq:FieldTransformations}.

The basic idea is to find a rotation $\Rotation{R} (\alpha, \beta,
\gamma)$ to maximize the quantity\footnote{For general systems in
  general orientations, relations like the usual $\wavefield^{\ell,
    -m} = (-1)^{\ell} \co{\wavefield}^{\ell, m}$ need not hold.  As a
  result, both terms in the sum over $m$ are required, to avoid mixing
  of the $\abs{m} \neq 2$ modes.}
\begin{equation}
  \label{eq:MaximizedQuantity}
  \begin{split}
    \capwavefield(\alpha, \beta, \gamma) &= \sum_{m=\pm 2}\,
    \abs{\Rotated{\wavefield}^{2,m}}^{2}
    \\
    &= \sum_{m=\pm 2}\, \abs{\sum_{m'=-2}^{2}\, \wavefield^{2,m'}\,
      \mathcal{D}^{(2)}_{m',m}(-\gamma, -\beta, -\alpha)}^{2}~,
  \end{split}
\end{equation}
where the $\mathcal{D}^{(2)}_{m',m}$ are given in terms of the Euler
angles by Eq.~\eqref{eq:WignerDMatrices}.  By considering the
relationship between the coordinate systems,
Eq.~\eqref{eq:AngleRelationship}, we can see that the radiation axis
is given by $\vec{a} = \Rotation{R} (\alpha, \beta, \gamma)\,
\hat{z}$.

O'Shaughnessy \etal~\cite{OShaughnessyEtAl:2011} introduced another
method, which finds an axis associated with the quadrupolar part of
the radiation field. They begin by defining
\begin{equation}
  \label{eq:OShaughnessyFormula}
  \langle L_{(a}\, L_{b)} \rangle = \sum_{\ell, m, m'}\,
  \co{\wavefield}^{\ell,m'}\, \braket{\ell, m'| L_{(a}\, L_{b)} |\ell,
    m}\, \wavefield^{\ell, m}~,
\end{equation}
where $L_{a}$ is the usual angular-momentum
operator~\cite{GoldbergEtAl:1967}, and for simplicity of presentation
we set $\int \abs{\wavefield}^{2}\, \d\Omega = 1$.  The radiation axis
is then defined as the dominant principal axis of this matrix---the
eigenvector with the eigenvalue of largest magnitude.  This problem
can be solved directly with standard algebraic techniques.

We find it useful to think of this method in a second way.  Basic
results from linear algebra show us that there exists a rotation
operator $\Rotation{R}$ such that the matrix $\Rotation{R} \left
  \langle L_{(a}\, L_{b)} \right \rangle \Rotation{R}^{-1}$ is
diagonal, and that the final column of this diagonalized matrix is the
dominant principal axis.  To put it another way, then, this method can
be regarded as finding a rotation operator that maximizes the $z$-$z$
component of the rotated matrix, in which case the radiation axis is
just $\vec{a} = \Rotation{R}\, \hat{z}$.

The similarity between the two methods becomes clear when we expand
this rotated matrix and take the $z$-$z$ component:
\begin{multline}
  \label{eq:OShaughnessyRotated}
  \left( \Rotation{R}\, \left \langle L_{(a}\, L_{b)} \right \rangle
    \Rotation{R}^{-1} \right)_{zz} \\
  \begin{aligned}[c]
    &= \sum_{\ell, m, m'}\, \co{\wavefield}^{\ell, m'}\, \braket{
      \ell,m' | \Rotation{R}\, L_{z}\, L_{z}\, \Rotation{R}^{-1} |
      \ell,m} \wavefield^{\ell, m}
    \\
    &= \sum_{\ell, m, m'}\, \Rotated{\co{\wavefield}}^{\ell, m'}\,
    \braket{ \ell,m' | L_{z}\, L_{z}\, | \ell,m}
    \Rotated{\wavefield}^{\ell, m}
    \\
    &= \sum_{\ell, m}\, m^{2}\, \abs{\Rotated{\wavefield}^{\ell,
        m}}^{2}~.
  \end{aligned}
\end{multline}
O'Shaughnessy \etal suggest the possibility of limiting the range of
the sum to just $\ell=2$.  If we further limit the sum to $m = \pm 2$,
we have $m^{2} = 4$ times the quantity $\capwavefield$ given in
Eq.~\eqref{eq:MaximizedQuantity}, and the method of
\cite{OShaughnessyEtAl:2011} reduces to that of
\cite{SchmidtEtAl:2011}.

Important differences remain between the implementations possible with
the two methods, however.  When O'Shaughnessy \etal sum over all
relevant $m$ modes, they are rotating the $\ell$ components of the
waveform, which are geometrically meaningful.  Thus, the full matrix
$\langle L_{(a} L_{b)} \rangle$ as they define it is a tensor and
therefore obeys standard rotation rules.  If we limit the sum over $m$
modes, we have a quantity that does \emph{not} behave properly under
rotations.  This difference means that $\Rotation{R}$ can be solved
for algebraically in the method of O'Shaughnessy \etal, while the
method of Schmidt \etal requires a more active maximization procedure.

Nonetheless, we note that the method of Schmidt \etal, if implemented
carefully, can be made quite accurate and efficient.  Because the
right-hand side of Eq.~\eqref{eq:MaximizedQuantity} (and even its
derivatives) can be easily expressed as a known analytic function of
the angles $\alpha$, $\beta$, $\gamma$, the problem is perfectly
suited to numerical optimization.  We find that it is very easy to
implement, with the code converging to the correct radiation axis
within roughly $\SI{e-8}{\radian}$, typically using fewer than 10
function evaluations.  This method is also quite robust, requiring no
initial guess for the radiation axis.  The speed and accuracy of this
code, then, are essentially the same as the speed and accuracy of
code implementing the method of O'Shaughnessy \etalp

\subsection{Degeneracies}
\label{sec:Degeneracies}

In the discussion above, we glossed over a pair of degeneracies
present in both of these methods.  The first is trivial: the radiation
axis produced by either method is really a directionless axis, rather
than the directed axis we have assumed.  Roughly speaking, this means
that $\vec{a}$ may be either parallel or anti-parallel to the orbital
angular momentum.  This degeneracy may be resolved by any convenient
means, such as comparison with the coordinate angular velocity.  In
the following, we assume that $\vec{a}$ is chosen to lie parallel to
the angular momentum or---at least---points in a consistent direction
from moment to moment.

The second degeneracy, however, exhibits a significant flaw in the
methods as presented: both are invariant under rotations about the
radiation axis, and therefore do not fix the rotation $\FrameRot(t)$
uniquely.  We can see this explicitly by looking at the behavior of
the modes under such a rotation:
\begin{equation}
  \label{eq:ModeTransformation_zAxis}
  \Rotated{\wavefield}^{\ell, m} \to \Rotated{\wavefield}^{\ell, m}\,
  \e^{\i\, m\, \gamma}~,
\end{equation}
where $\gamma$ is the angle of the rotation.  Using this in either
Eq.~\eqref{eq:MaximizedQuantity} or
Eq.~\eqref{eq:OShaughnessyFormula}, we see that the phase factor
cancels out, leaving no change to the expressions.  We need to impose
another condition to make these problems well posed.  Both Schmidt
\etal and O'Shaughnessy \etal break the degeneracy by simply setting
the final Euler angle of the rotation to 0.  In our conventions, this
means setting $\gamma=0$ at all times.  But this choice means that the
rotation $\Rotation{R}(\alpha, \beta, \gamma)$ depends on the inertial
frame with respect to which the Euler angles are defined.  For general
precessing systems, it will affect the phase of the final waveform in
highly nontrivial ways, as we demonstrate in
Sec.~\ref{sec:EffectsOnWaveforms}.

Nonetheless, the choice of $\gamma=0$ does make the particular problem
of finding the radiation axis well posed.  In the next section we take
that radiation axis, and use the freedom in $\gamma$ to construct a
geometrically meaningful frame.  This requires abandoning locality in
time: while the methods of Refs.~\cite{SchmidtEtAl:2011}
and~\cite{OShaughnessyEtAl:2011} can be applied for each $t$
separately, our method will result in an ordinary differential
equation for the rotation matrix.

\section{Minimizing rotation}
\label{s:FindingMinimalRotFrame}

The techniques just described give us one particular rotation
$\YawfulRot(t) = \Rotation{R} \big(\alpha(t), \beta(t), 0\big)$ that
aligns the inertial frame with the radiation axis.  But the previously
noted freedom in $\gamma$ means that we can first perform a rotation
$\GammaRot(t)$ by an angle $\gamma$ about the $z$ axis without
affecting the radiation axis.  We now construct a rotation
\begin{equation}
  \label{eq:TotalRotation}
  \Rotation{R}(t) = \YawfulRot(t)\, \GammaRot(t)~,
\end{equation}
and solve for $\GammaRot(t)$ such that the minimal-rotation condition,
Eq.~\eqref{eq:MinimalRotationCondition}, is satisfied.  The new
rotation $\Rotation{R}(t)$ will simultaneously satisfy the
minimal-rotation condition and align the inertial $z$ axis with the
radiation axis.  To find $\GammaRot$, we express
Eq.~\eqref{eq:MinimalRotationCondition} in terms of the rotation
operator $\Rotation{R}$ alone, making use of generators in the Lie
algebra $\so3$, and various relations noted in
Appendix~\ref{sec:Conventions}.  We then apply this to the case where
$\Rotation{R}$ is decomposed as in Eq.~\eqref{eq:TotalRotation},
allowing us to solve for the minimal rotation.

\subsection{The minimal-rotation condition in terms of the rotation
  operator}
\label{sec:MinimalRotationOperator}
We begin by defining the equivalents of the instantaneous rotation
axis $\FrameRot$ and the radiation axis $\vec{a}$ using the
isomorphism $\sigma$ which maps 3-vectors into $\so3$, given in a
Cartesian basis by Eq.~\eqref{eq:VectorsIsomorphism}.  We write
$\FrameRotGen = \sigma(\FrameRot)$ and $\Generator{A} =
\sigma(\vec{a})$.  Now, the dot product can also be defined for
elements of $\so3$ [as $-1/2$ times the trace of the product
matrix; see Eq.~\eqref{eq:DotProduct}], allowing us to rewrite the
minimal-rotation condition as
\begin{equation}
  \label{eq:MinimalRotationCondition_so3}
  \FrameRotGen \cdot \Generator{A}=0~.
\end{equation}
Here, $\FrameRotGen$ is unknown, and $\Generator{A}$ is time
dependent.  Therefore, we now translate these into expressions in
terms of the rotation operator and the basis element $\Generator{Z} =
\sigma(\hat{z})$.

The formula for $\Generator{A}$ is simple.  Recall from
Sec.~\ref{s:LocatingRadiationAxis} that $\vec{a} = \Rotation{R}\,
\hat{z}$.  In terms of generators, this is $\Generator{A} =
\Rotation{R}\, \Generator{Z}\, \Rotation{R}^{-1}$.  The formula for
$\FrameRotGen$ can be found by considering
Eq.~\eqref{eq:StationaryToRotatingVectorDeriv}, applied to \emph{any}
vector $\vec{v}$ that is stationary in the rotating frame.  If we
define $\vec{v}_{0} \coloneqq \Rotation{R}^{-1}(0)\, \vec{v}(0)$, we
can write $\vec{v}(t) = \Rotation{R}(t)\, \vec{v}_{0}$ in the inertial
frame.  In $\so3$, this is written $\Generator{V} = \Rotation{R}\,
\Generator{V}_{0}\, \Rotation{R}^{-1}$.  Then
Eq.~\eqref{eq:StationaryToRotatingVectorDeriv} becomes
\begin{align}
  \label{eq:StationaryToRotatingVectorDeriv_Operators}
  \frac{\d}{\d t}\, \big(\Rotation{R}\, \Generator{V}_{0}\,
  \Rotation{R}^{-1} \big) &= [\FrameRotGen, \Rotation{R}\,
  \Generator{V}_{0}\, \Rotation{R}^{-1}] \\ &= [ \dot{\Rotation{R}}\,
  \Rotation{R}^{-1}, \Rotation{R}\, \Generator{V}_{0}\,
  \Rotation{R}^{-1} ]~,
\end{align}
where the second line comes from expanding the derivative using
Eq.~\eqref{eq:DerivativeOfRotation}.  If equality is to hold for
arbitrary $\vec{v}$, we must have
\begin{equation}
  \label{eq:FrameRotationGenerator}
  \FrameRotGen = \dot{\Rotation{R}}\, \Rotation{R}^{-1}~.
\end{equation}

Now, using these expressions for $\Generator{A}$ and $\FrameRotGen$ in
Eq.~\eqref{eq:MinimalRotationCondition_so3} and rearranging a little,
we get another form of the minimal-rotation condition:
\begin{equation}
  \label{eq:MinimalRotationCondition3}
  \big( \dot{\Rotation{R}}\, \Rotation{R}^{-1} \big) \cdot \big(
  \Rotation{R}\, \Generator{Z}\, \Rotation{R}^{-1} \big) = 0~.
\end{equation}
This is precisely the minimal-rotation condition of
Eq.~\eqref{eq:MinimalRotationCondition} in operator form.  We can
simplify this expression slightly.  Noting that the dot product is
invariant under rotations, we apply the inverse rotation to each part
of the product, obtaining
\begin{equation}
  \label{eq:MinimalRotationCondition4}
  \big( \Rotation{R}^{-1}\, \dot{\Rotation{R}} \big) \cdot
  \Generator{Z} = 0~.
\end{equation}

\subsection{Solving for the initial rotation}
\label{sec:SolvingForInitialRotation}
To find $\Rotation{R}$ satisfying the minimal-rotation condition, we
now insert Eq.~\eqref{eq:TotalRotation} into
Eq.~\eqref{eq:MinimalRotationCondition4}.  Assuming that $\YawfulRot$
is known, and using $\GammaRot = \exp(\gamma\, \Generator{Z})$, this
will give us a condition on $\gamma$.  First, we calculate
\begin{equation}
  \label{eq:RinverseRdot}
  \Rotation{R}^{-1}\, \dot{\Rotation{R}} = \e^{-\gamma\,
    \Generator{Z}}\, \YawfulRot^{-1}\, \YawfulRotDot\, \e^{\gamma\,
    \Generator{Z}} + \dot{\gamma}\, \Generator{Z}~.
\end{equation}
Note that conjugation by $\e^{-\gamma\, \Generator{Z}}$ does not
affect the component along $\Generator{Z}$.  Therefore, plugging this
result into Eq.~\eqref{eq:MinimalRotationCondition4} and rearranging,
we obtain
\begin{equation}
  \label{eq:MinimalRotation_General}
  \dot{\gamma} = \big( -\YawfulRot^{-1}\, \YawfulRotDot \big) \cdot
  \Generator{Z}~.
\end{equation}
Because $\YawfulRot$ is known, we can simply evaluate the right-hand
side, integrate to find $\gamma(t)$, and insert this back into
Eq.~\eqref{eq:TotalRotation} to find a minimal-rotation operator that
takes the $z$ axis into the radiation axis.

We emphasize that the derivations of
Eqs.~\eqref{eq:MinimalRotationCondition4}
and~\eqref{eq:MinimalRotation_General} did not assume any features of
$\Rotation{R}$ and $\YawfulRot$ other than the fact that they rotate
the $z$ axis of the inertial frame into the radiation axis.  In
particular, we did not assume that the final Euler angle was zero---or
indeed use any expression in terms of the Euler angles.  Also, though
inspired by the Euler angles, the definition $\GammaRot \coloneqq
\exp(\gamma\, \Generator{Z})$ is independent of coordinates on
$\SO3$.\footnote{Note the distinction between the choice of
  \emph{basis} for $\mathbb{V}^{3}$ and the choice of
  \emph{coordinates} for $\SO3$.  Here, the basis $(\hat{x}, \hat{y},
  \hat{z})$ produces the canonical basis of generators
  $(\Generator{X}, \Generator{Y}, \Generator{Z})$ for $\so3$ by
  eigenvector problems.  In particular, $\Generator{Z}$ represents the
  unique generator having eigenvector $\hat{z}$ with eigenvalue 0, and
  eigenvector $(\hat{x} + \i\, \hat{y}) / \sqrt{2}$ with eigenvalue
  $\i$.  Also, the exponential function is defined geometrically (see
  Eq.~\eqref{eq:RotationGeneratorRelationship} or
  Ref.~\cite{DuistermaatKolk:1999}), which shows that $\exp(\gamma\,
  \Generator{Z})$ is independent of coordinates on $\SO3$.  Similarly,
  the arbitrary choice of $z$ axis will not affect the form of these
  equations; we can choose any unit vector as the $z$ axis, and as
  long as the rotation operators take that vector into the radiation
  axis, the expressions will not change.} %
Therefore, Eq.~\eqref{eq:MinimalRotation_General} is a geometrically
general equation: it should hold regardless of any coordinates we
might choose for $\SO3$, and should take the same form regardless of
the inertial frame we use to find the radiation axis.  One result of
this is the fact that frames satisfying the minimal-rotation condition
are unique up to a constant overall rotation about the radiation axis,
corresponding to the integration constant obtained from integrating
Eq.~\eqref{eq:MinimalRotation_General}.  We discuss this freedom
further in Sec.~\ref{sec:HybridWaveforms}.

Nonetheless, for the purposes of implementation, an explicit formula
involving the Euler angles will be useful.  When $\YawfulRot(t) =
\Rotation{R} \big(\alpha(t), \beta(t), 0\big)$, a straightforward
calculation using Eqs.~\eqref{eq:DerivativeOfRotationAsGenerator}
and~\eqref{eq:GeneratorConjugation} gives us\footnote{Near the
  coordinate singularities at $\beta = 0$ and $\beta=\pi$, the value
  of $\alpha$ will contain substantial numerical noise.
  Differentiating $\alpha$, as in this equation, simply magnifies that
  noise.  For some configurations, we find improved numerical results
  when integrating this equation by parts and implementing it as
  $\gamma = -\alpha\, \cos\beta - \int\, \alpha\, \dot{\beta}\, \sin
  \beta\, \d t$.}
\begin{equation}
  \label{eq:MinimalRotation_Euler}
  \dot{\gamma} = - \dot{\alpha}\, \cos\beta~.
\end{equation}
Schmidt \etal~\cite{SchmidtEtAl:2011} pointed out that the orbital
frequency in the rotating frame given by $\YawfulRot$ should be
roughly $\Omega + \dot{\alpha}\, \cos\beta$ (in our notation), where
$\Omega$ is the magnitude of the orbital frequency measured in the
non-rotating frame.  Thus our adjustment to their technique can be
thought of as removing that second term, so that the orbital frequency
in the rotating frame given by $\Rotation{R}$ should be roughly
$\Omega$.  We will return to this observation in
Sec.~\ref{sec:PostNewtonianWaveforms} to discuss a simplification of
the post-Newtonian representation of precessing systems.

\section{The effects of enforcing the minimal-rotation condition}
\label{sec:EffectsOfMinimalRotation}
No preferred inertial frame exists for general precessing systems.
Indeed, no preferred \emph{axis} exists for choosing an inertial
frame.  When the total angular momentum $\vec{J}$ points in a constant
direction, this can be a useful choice of axis, but $\vec{J}$ will
change direction for inspiralling precessing systems.  We might
therefore expect data from different numerical simulations, for
example, to be presented in different inertial frames.  A key concern,
then, is the behavior of the rotated frame under fixed rotations of
the inertial frame.  We now demonstrate that a frame aligned with the
radiation axis using $\gamma = 0$ behaves poorly under such rotations,
whereas such a frame with the minimal-rotation condition imposed is
essentially invariant.  First, we observe the motion of the rotated
axes in the two cases.  Then, we inspect the behavior of the phase of
the waveform decomposed in the rotating frames.

\subsection{Rotation history in different inertial frames}
\label{sec:RotationHistory}
\begin{figure}
  \includegraphics[width=.23\columnwidth, bb=20 0 210 210,
  clip]{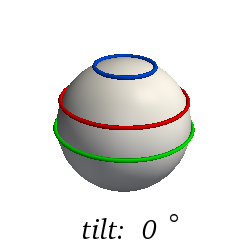}
  \includegraphics[width=.23\columnwidth, bb=20 0 210 210,
  clip]{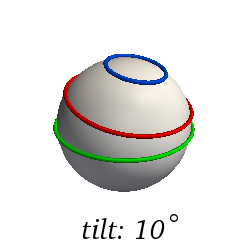}
  \includegraphics[width=.23\columnwidth, bb=20 0 210 210,
  clip]{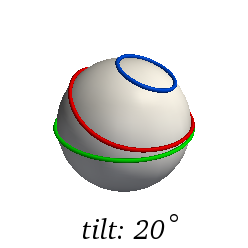}
  \includegraphics[width=.23\columnwidth, bb=20 0 210 210,
  clip]{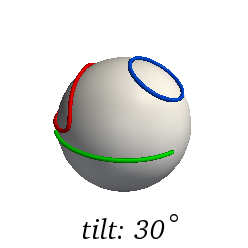}\\
  \includegraphics[width=.23\columnwidth, bb=20 0 210 210,
  clip]{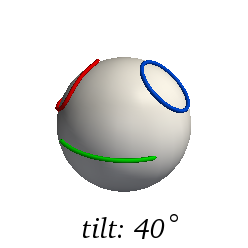}
  \includegraphics[width=.23\columnwidth, bb=20 0 210 210,
  clip]{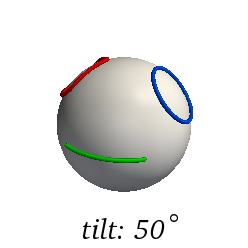}
  \includegraphics[width=.23\columnwidth, bb=20 0 210 210,
  clip]{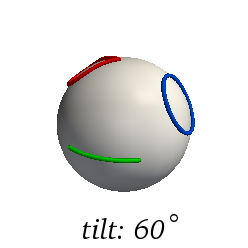}
  \includegraphics[width=.23\columnwidth, bb=20 0 210 210,
  clip]{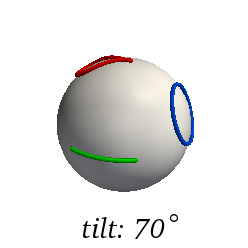}\\
  \includegraphics[width=.23\columnwidth, bb=20 0 210 210,
  clip]{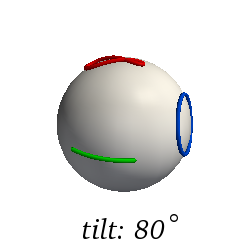}
  \includegraphics[width=.23\columnwidth, bb=20 0 210 210,
  clip]{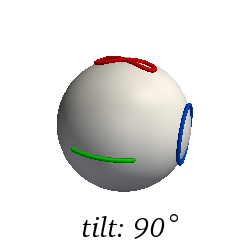}
  \caption{ \label{fig:SchmidtRotations} %
    Rotation histories for a simple precessing system when $\gamma =
    0$.  The axis denoted by the blue curve (the small, perfectly
    circular loop) precesses through a single cycle around a fixed
    axis that is tilted by varying amounts.  The other axes of the
    tracked frame (red and green curves) take paths that depend on the
    inclination of the axis around which the precession occurs,
    relative to the inertial frame.  The $y$ axis, in particular, is
    forced to remain on the $x$-$y$ plane of the visualization
    coordinates, by the choice of $\gamma = 0$.  If we had defined
    Euler angles by $z$-$x$-$z$ rotations, as in
    Ref.~\cite{SchmidtEtAl:2011}, then it would have been the $x$ axis
    that was forced to remain on this plane.  %
  }
\end{figure}

For any rotation $\Rotation{R}(t)$, we visualize its ``rotation
history'' by plotting the paths of the tips of the rotated basis
vectors, $\hat{x}'$, $\hat{y}'$, and $\hat{z}'$, on the unit sphere.
We demonstrate the rotation histories for a toy model in which the
radiation axis $\vec{a}$ precesses about a fixed axis $\vec{F}$, where
$\vec{a}$ is inclined to $\vec{F}$ by an angle of $\ang{25}$.  In the
first panel of Fig.~\ref{fig:SchmidtRotations}, we show the rotation
history for this system when $\vec{F}$ is along the $z$ axis and the
final Euler angle is simply set to $\gamma=0$.  The blue circle traces
the path of the radiation axis---which coincides with the $\hat{z}'$
basis vector---while the red and green curves trace the paths of the
$\hat{x}'$ and $\hat{y}'$ axes, respectively.

If, instead of being aligned with the $z$ axis of the inertial frame,
the $\vec{F}$ axis is tipped, we obtain different rotation histories.
Later panels of Fig.~\ref{fig:SchmidtRotations} show the histories
when $\vec{F}$ is tilted by the given amount.  In each panel, the blue
curve remains the same, being simply shifted on the sphere by the
given tilt.  However, the paths of the other two axes of the adapted
frame change drastically as the inclination of $\vec{F}$ is changed,
even undergoing topological transitions as this inclination passes the
$\ang{25}$ precession inclination and its $\ang{65}$ complement.  The
tracked frame thus has a time-dependent rotation about the radiation
axis, which will show up as a time-dependent modulation of the
waveform phase.  This modulation is determined by the choice of
inertial frame, and therefore the phase of a waveform measured in a
frame obtained with $\gamma=0$ is not invariant in any useful sense.
This phase modulation will be examined directly in
Sec.~\ref{sec:EffectsOnWaveforms}.

\begin{figure}
  \includegraphics[width=.23\columnwidth, bb=20 0 210 210,
  clip]{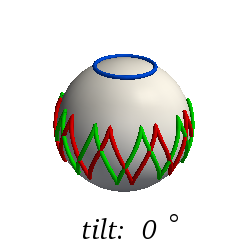}
  \includegraphics[width=.23\columnwidth, bb=20 0 210 210,
  clip]{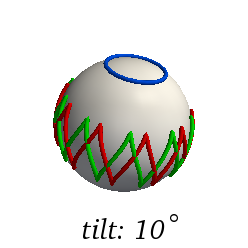}
  \includegraphics[width=.23\columnwidth, bb=20 0 210 210,
  clip]{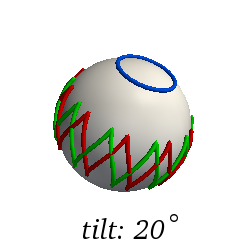}
  \includegraphics[width=.23\columnwidth, bb=20 0 210 210,
  clip]{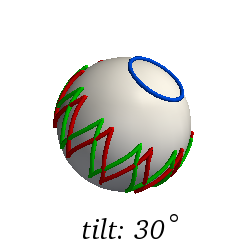}\\
  \includegraphics[width=.23\columnwidth, bb=20 0 210 210,
  clip]{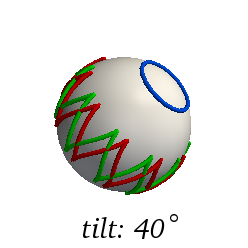}
  \includegraphics[width=.23\columnwidth, bb=20 0 210 210,
  clip]{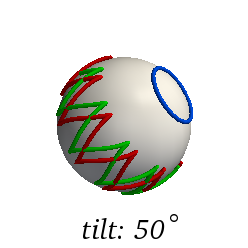}
  \includegraphics[width=.23\columnwidth, bb=20 0 210 210,
  clip]{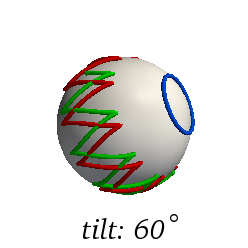}
  \includegraphics[width=.23\columnwidth, bb=20 0 210 210,
  clip]{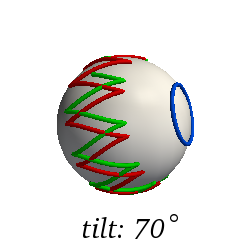}\\
  \includegraphics[width=.23\columnwidth, bb=20 0 210 210,
  clip]{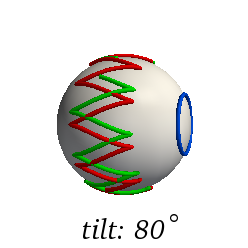}
  \includegraphics[width=.23\columnwidth, bb=20 0 210 210,
  clip]{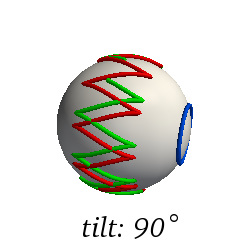}
  \caption{ \label{fig:OurRotations}%
    Rotation histories for a simple precessing system when $\gamma$ is
    set by Eq.~\eqref{eq:MinimalRotation_Euler}.  These systems are
    just the same as in Fig.~\ref{fig:SchmidtRotations}, except that
    the frame is given a final rotation about the $z$ axis to satisfy
    the minimal-rotation condition.  The paths are identical in each
    case, but tilted with respect to each other.  This shows that the
    minimally rotated frame is essentially invariant under fixed
    rotations of the inertial frame.  Note that the paths in these
    figures represent many precession cycles, each corresponding to a
    single cycle of the ``zigzag'' pattern.  }
\end{figure}

We can repeat this comparison of rotation histories when the third
Euler angle is set by the minimal-rotation condition,
Eq.~\eqref{eq:MinimalRotation_Euler}.  The results are shown in
Fig.~\ref{fig:OurRotations}.  In this case, the rotation histories
have the same shapes, but are simply tilted with respect to each
other.  That
similarity shows that frames constructed with the minimal-rotation
condition are essentially\footnote{There is still an overall freedom
  in a fixed rotation about the initial radiation axis.  This
  corresponds to the standard ambiguity in orientation, which must be
  fixed by other methods, discussed in
  Sec.~\ref{sec:HybridWaveforms}.} invariant under fixed rotations of
the inertial frame.  As we will now see, the waveform measured in such
a frame is similarly invariant.

\subsection{Waveforms in different inertial frames}
\label{sec:EffectsOnWaveforms}
The methods set forth in Refs.~\cite{SchmidtEtAl:2011}
and~\cite{OShaughnessyEtAl:2011} fix the axis of the rotated frame in
an invariant way, but rely on coordinate-degrees of freedom to fix
rotations about that axis.  Such a rotation translates directly into
the phase of the waveform measured in the rotated frame.  Since the
minimal-rotation condition is imposed with a rotation about this
radiation axis, only the phase of the waveform is affected.
Therefore, we ignore the waveform amplitude, as it is invariant as
soon as the radiation axis is fixed.  We examine the waveform phase
with two examples---a simple analytical example first, followed by a
more realistic model.

Imagine a system with very small precession, where the radiation axis
moves with constant speed along a vanishingly narrow cone centered
about the $z$ axis.  Because the cone is so narrow, we might hope that
the phase of the waveform $\wavefield$ measured in the inertial frame
will be close to the phase measured in the rotated frame.  Assuming
the radiation axis is in the $x$-$z$ plane at time $t=0$, the rotation
found by the methods of Refs.~\cite{SchmidtEtAl:2011}
and~\cite{OShaughnessyEtAl:2011} will be $\Rotation{R}(\dot{\alpha}\,
t, \beta, 0)$, for constants $\dot{\alpha}$ and $\beta$.  We can
relate the modes of the waveform measured in the rotated frame
$\Rotated{\wavefield}^{\ell, m}$ to the modes measured in the inertial
frame $\wavefield^{\ell, m}$ using Eq.~\eqref{eq:FieldTransformation}.
If we approximate $\beta \approx 0$, then the Wigner matrices
$\mathcal{D}^{(\ell)}_{m,m'}$ are nonzero only for $m=m'$, and we have
\begin{subequations}
  \label{eq:SimpleExample}
  \begin{equation}
    \label{eq:ModesInSimpleExample}
    \Rotated{\wavefield}^{\ell, m} \approx \wavefield^{\ell, m}\,
    \e^{-\i\, m\, \dot{\alpha}\, t}~.
  \end{equation}
  That is, the waveform acquires an additional linearly increasing
  phase in the rotated frame.  If the complex phase of this mode is
  $\phi^{\ell, m}(t)$, the change in going from a non-precessing
  system to a system precessing with a very small opening angle is
  \begin{equation}
    \label{eq:PhaseChange1}
    \phi^{\ell, m}(t) \to \phi^{\ell, m}(t) - m\, \dot{\alpha}\, t~.
  \end{equation}
\end{subequations}
Note that the additional phase only depends on the number of times the
system has precessed, not the size of the precession angle.  If, on
the other hand, we impose the minimal-rotation condition,
Eq.~\eqref{eq:MinimalRotation_Euler} gives us $\gamma \approx
-\dot{\alpha}\, t$, which cancels the additional phase, so that
$\Rotated{\wavefield}^{\ell, m} \approx \wavefield^{\ell, m}$.  In
this sense, the minimal-rotation frame is much more natural.

More importantly, however, a frame with minimal rotation behaves
nicely under fixed rotations of the inertial frame.  For example, we
take the same system as above, but tilt the inertial frame slightly so
that the precession cone lies close to, but does \emph{not} contain
the $z$ axis.  In this case, the rotation will be
$\Rotation{R}(\varphi, \vartheta, 0)$, where $(\vartheta, \varphi)$
are the usual spherical coordinates of the axis, which we can
calculate by simple trigonometry.  Using this result to transform the
modes, we find
\begin{subequations}
  \label{eq:SimpleExample_Tilted}
  \begin{equation}
    \label{eq:ModesInSimpleExampleTilted}
    \Rotated{\wavefield}^{\ell, m} \approx \wavefield^{\ell, m}\,
    \e^{-\i\, m\, \arccot [\cot(\dot{\alpha}\, t) + \theta
      \csc(\dot{\alpha}\, t)]}~,
  \end{equation}
  where $\dot{\alpha}$ is the same constant as above and $\theta>1$ is
  a constant that depends on the particular values of the precession
  and the tilt angles.\footnote{In fact, this form of the equation
    applies for all small precession angles and tilts.  If $B$ is the
    tilt angle, then $\theta = B/\beta$, so $\theta=0$ corresponds to
    the case where the precession cone is centered on the $z$ axis,
    and Eq.~\eqref{eq:SimpleExample_Tilted} reduces to
    Eq.~\eqref{eq:SimpleExample}.  Similarly, $0 < \theta < 1$
    corresponds to a small tilt, for which the $z$ axis is still
    within the precession cone.  In this case, the additional phase is
    roughly linear, with a superposed oscillation.  $\theta=1$
    corresponds to the case where the radiation axis passes through
    the $z$ axis, which is the coordinate singularity of the Euler
    angles, meaning that the waveform phase in the $\gamma=0$ frame is
    actually undefined.  Finally, $\theta>1$ corresponds to a tilt
    that is larger than the precession angle, so that the $z$ axis is
    not enclosed in the precession cone.}  In this case, the change to
  the waveform phase is bounded, but oscillatory:
  \begin{equation}
    \label{eq:PhaseChange2}
    \phi^{\ell, m}(t) \to \phi^{\ell, m}(t) - m\, \arccot
    [\cot(\dot{\alpha}\, t) + \theta \csc(\dot{\alpha}\, t)]~.
  \end{equation}
\end{subequations}
Thus, a slight change in the inertial frame causes a drastic change in
the behavior of the waveform modes, which is associated with the
topological change seen in Fig.~\ref{fig:SchmidtRotations} when the
tilt exceeds \ang{25}.  In the minimal-rotation frame, on the other
hand, the phase change in Eq.~\eqref{eq:ModesInSimpleExampleTilted} is
counteracted by the adjustment to $\gamma$, and we still have
$\Rotated{\wavefield}^{\ell, m} \approx \wavefield^{\ell, m}$.

\begin{figure}
  \includegraphics{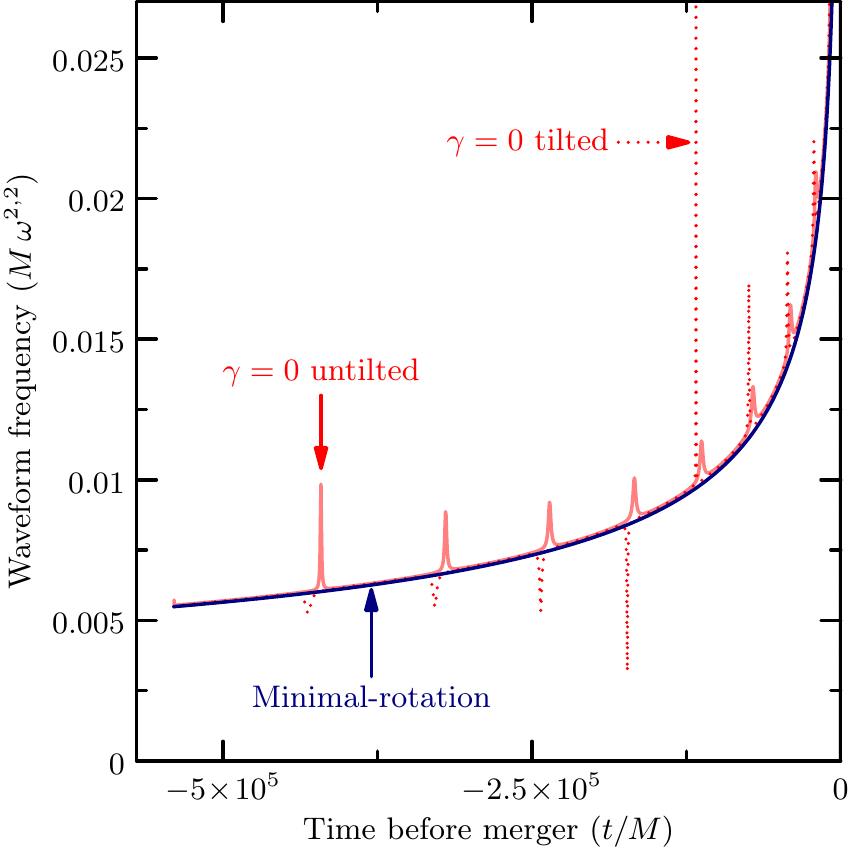}
  \caption{ \label{fig:WaveformFrequencies}%
    Waveform frequency in various frames aligned with the radiation
    axis.  Here, waveform frequency refers to the time derivative of
    the phase of the $(\ell, m) = (2,2)$ mode of the gravitational
    waveform.  The solid red line shows the frequency measured in a
    frame derived from the inertial frame by a rotation in which the
    third Euler angle $\gamma$ is set to 0, while the solid blue line
    shows the same quantity in a frame for which $\gamma$ satisfies
    the minimal-rotation condition.  Clearly, the latter curve is much
    smoother.  We also show as dotted lines the same quantities when
    the physical system is tilted by \ang{10}.  The dotted blue line
    coincides with the solid blue line, showing the invariance of the
    waveform in that frame.  %
  }
\end{figure}

These features also show up in systems with significant precession.
We now turn to a post-Newtonian model of such a binary.  The system we
choose has equal-mass black holes, with spins $\chi=0.99$ initially
parallel to each other and orthogonal to the orbital angular momentum,
which initially coincides with the $z$ axis.  The precession cone has
an opening angle of roughly \ang{15} initially, gradually widening to
about \ang{21}.  Using the post-Newtonian waveform, we can the find
the radiation axis with the methods described in
Sec.~\ref{s:LocatingRadiationAxis}, then decompose the modes of the
waveform either in a frame with $\gamma=0$ or in a frame with minimal
rotation.

Again, we see the two features identified above.  First, the waveform
decomposed in the minimal-rotation frame appears to be smoother than
the waveform decomposed in the $\gamma=0$ frame.  In particular, while
the amplitudes are identical in the two frames, the phase of the
$(\ell, m) = (2,2)$ mode in the $\gamma=0$ frame is constantly
increasing relative to the phase in the minimal-rotation frame, and
jumps each time the radiation axis passes near the $z$ axis (each time
$\dot{\alpha}\, \cos\beta$ is large).  We plot the \emph{frequency} of
the $(2,2)$ mode measured in the two frames in
Fig.~\ref{fig:WaveformFrequencies}, where the phase jumps show up as
spikes.

\begin{figure}
  \includegraphics{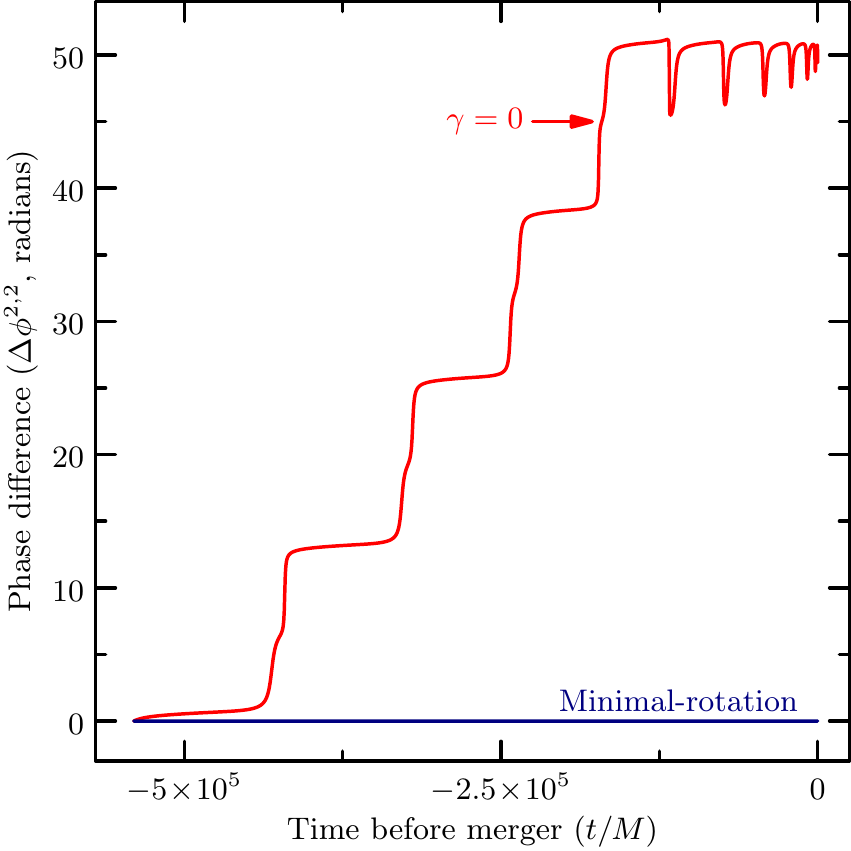}
  \caption{ \label{fig:WaveformPhases} %
    Change of phase measured in frames aligned to the radiation axis
    when the physical system is tilted by \ang{10}.  This phase
    difference is defined by Eq.~\eqref{eq:PhaseDifference}, and
    plotted for two cases: the first (in red) where $\phi^{2,2}$
    refers to phases measured in a frame obtained from the inertial
    frame with a rotation in which the final Euler angle $\gamma$ is
    set to 0; the second (in blue) where $\phi^{2,2}$ refers to phases
    measured in frames satisfying the minimal-rotation condition.  The
    change in phase for the minimal-rotation frames is 0 to within
    numerical error (roughly a part in $10^{5}$ here).  Note that this
    waveform extends for roughly the length of time such a system
    would be in the sensitive band of Advanced LIGO if the total mass
    were about $10\,\MSun$.  %
  }
\end{figure}

Second, the waveform phase in the minimal-rotation frame is invariant
(up to a constant) under overall rotations of the inertial frame in
which the waveform is measured---which is not the case for the
$\gamma=0$ frame.  We illustrate this by tilting the post-Newtonian
system by a \ang{10} rotation about the $y$ axis, and redoing the
decomposition in the two frames.  The frequencies for this rotated
system are also plotted (as dotted lines) in
Fig.~\ref{fig:WaveformFrequencies}, where we see that the curves for
the minimal-rotation frame lie on top of each other, while the
$\gamma=0$ curve changes but still exhibits spikes.

Figure~\ref{fig:WaveformPhases} plots the phase-differences of the
(2,2) mode between the two post-Newtonian evolutions differing by a
rotation by $\ang{10}$,
\begin{equation}
  \label{eq:PhaseDifference}
  \Delta \phi^{2,2} = \phi^{2,2}_{\text{untilted}} -
  \phi^{2,2}_{\text{tilted}}.
\end{equation}
The minimal-rotation frame is invariant under this rotation, and
indeed the phases are identical to within the numerical error (as
measured by convergence of the integration of
Eq.~\eqref{eq:MinimalRotation_Euler}).  The coordinate-dependent
choice $\gamma=0$, however, results in phase differences of multiple
gravitational-wave cycles.

While we designed this example to be a rigorous test of the
frame-alignment techniques, the later stages of inspiral and merger
provide an even more stringent test, as precession frequency scales
with orbital velocity.  The minimal-rotation condition provides an
easily implemented solution to the problems presented by the
$\gamma=0$ frame.


\section{Further applications of the minimal-rotation frame}
\label{sec:Applications}

The minimal-rotation frame aligned with the radiation axis allows us
to describe gravitational waveforms from precessing binaries very
nicely.  The amplitude and phase are smoothly varying functions, and
are invariant (up to an overall phase offset) under rotations of the
inertial frame.  The significance of this construction is broader than
one might immediately imagine.  We now discuss some of the most
important uses of gravitational waveforms, and show that the
minimal-rotation frame can make the necessary manipulations much
simpler.

\subsection{Post-Newtonian waveforms}
\label{sec:PostNewtonianWaveforms}

Buonanno, Chen, and Vallisneri~\cite{BuonannoEtAl:2003} proposed the
construction of template waveforms for precessing binaries, expressing
the metric perturbation using the minimal-rotation frame (which they
simply called the ``precessing frame'').  Their motivation was to
separate the response of a detector into two parts: one describing the
intrinsic qualities of the waveform; the other describing the position
and orientation of the detector.

This description differs from the more usual
approach~\cite{ApostolatosEtAl:1994, Kidder:1995, ArunEtAl:2009,
  CampanelliEtAl:2009} of tracking the binary's orbital parameters in
a frame related to the inertial frame by a rotation taking the $z$
axis into the Newtonian angular momentum $\LN$, with the final Euler
angle set to $\gamma=0$.  This results in a nontrivial relationship
between the orbital frequency in the inertial frame,
$\Omega_{\text{inertial}}$, and the orbital frequency in the
instantaneous orbital plane, $\Omega_{\text{orbital}}$:
\begin{equation}
  \label{eq:InertialOrbital}
  \Omega_{\text{inertial}} = \Omega_{\text{orbital}} + \dot{\alpha}\,
  \cos\beta~.
\end{equation}
In such a framework, this equation must be integrated as part of the
post-Newtonian system, along with the equations for the angular
momentum, spins, and the usual flux and orbital energy, resulting in
the motion of the binary as a function of time.  This motion is then
inserted into expressions for the gravitational waveform in the
inertial frame.  Because of the complicated motion, these expressions
are necessarily even more elaborate than the standard expressions for
motion in the orbital plane~\cite{ArunEtAl:2009}.

We can substantially simplify the analytical prescription by evolving
the orbital elements in the inertial frame, but writing the waveform
in terms of spin-weighted spherical harmonics in the rotating frame
aligned with the angular momentum~\cite{ApostolatosEtAl:1994}.  This
is particularly easy if the frame satisfies the minimal-rotation
condition, Eq.~\eqref{eq:MinimalRotationCondition}.  In this case, the
instantaneous rotation vector is
\begin{equation}
  \label{eq:FrameRotLN}
  \FrameRot = \LN \times \LNdot~.
\end{equation}
Assuming that $\LN$ is given by the angles $\alpha$ and $\beta$, then
as in Sec.~\ref{sec:SolvingForInitialRotation} we can define the
rotating frame by the operator
\begin{equation}
  \label{eq:PNFrame}
  \Rotation{R}\left( \alpha(t), \beta(t), -\textstyle{\int}\,
    \dot{\alpha}\, \cos\beta\, \d t \right)~.
\end{equation}
In such a frame, Eq.~\eqref{eq:InertialOrbital} becomes
\begin{equation}
  \label{eq:InertialOrbital2}
  \Omega_{\text{inertial}} = \Omega_{\text{orbital}}~.
\end{equation}
This means that the \pN orbital elements can be integrated in the
usual way.  That is, the orbital phase $\Phi$ obtained by integrating
the angular frequency is just the phase in the rotating frame.  This
can then be inserted into the standard
expressions~\cite{BlanchetEtAl:2008, WillWiseman:1996, AjithEtAl:2007,
  ArunEtAl:2009} for the waveform modes,\footnote{This prescription is
  sufficient at the level of knowledge of post-Newtonian waveforms
  given in the references.  However, by treating the waveform as if it
  were determined only by instantaneous positions and velocities, it
  neglects contributions which may be relevant at higher orders.  For
  example, the standard expressions for the waveform amplitudes assume
  accelerations are orthogonal to the orbital angular velocity, which
  is not the case for precessing systems.  Nonetheless, these
  contributions can be calculated and projected into the rotating
  frame.  As these are higher-order spinning terms, we ignore them at
  this level.}  which gives the waveform in the rotating frame.
Together with the rotation operator given by Eq.~\eqref{eq:PNFrame},
this describes the waveform completely.  In particular, it can readily
be transformed to the inertial frame using
Eq.~\eqref{eq:FieldTransformations}.

\subsection{Comparing waveforms and constructing hybrids}
\label{sec:HybridWaveforms}

Having obtained some model waveform, one of the first things we
typically do is compare the result to some other waveform.  For
example, we might run a numerical simulation at two different
resolutions or using two different numerical codes, and compare the
output waveforms to get an estimate for their accuracy.  Or we might
construct an analytical model, and compare it to a numerical model of
the same system.  And of course, numerical simulations can only
describe the last portion of the inspiral of a binary system, so we
frequently combine analytical and numerical waveforms into ``hybrid''
waveforms.  In each of these cases, the first step is to ensure that
the two waveforms are expressed in the same coordinate system, which
is generally referred to as \textit{alignment}.

With various simplifying assumptions, alignment boils down to setting
the time offset and the relative orientation of the frames in which
the waveforms are measured.  For a non-precessing binary, as discussed
in Sec.~\ref{sec:Introduction}, this further reduces to setting the
time offset and a single rotation around the $z$ axis.  That is, one
waveform is adjusted as
\begin{equation}
  \label{eq:WaveformTransformation}
  \h^{\ell, m} (t) \to \h^{\ell, m} (t + \Delta t)\, \e^{-\i\, m\,
    \Delta \Phi}~,
\end{equation}
for time offset $\Delta t$ and phase offset $\Delta \Phi$.  The
criteria used to choose those offsets typically ensure that the phase
and frequency of $\h^{2,2}$ are the same in both waveforms at a
particular instant, for example.\footnote{Many other possible criteria
  exist for choosing these offsets (see, e.g.,
  Ref.~\cite{Boyle:2011dy}) but the effect on the waveform should
  always be given by Eq.~\eqref{eq:WaveformTransformation}.  Therefore
  we ignore the particular criteria in use.}

Precession complicates this simple picture, because we can no longer
rotate the second system by just one angle $\Delta \Phi$.  In general,
the inertial frames will be related by some rotation $\RFrame$, which
rotates the second frame into the first.  We would need to solve for
this rotation, which might involve rotating the entire $\ell=2$
component of the waveform to minimize some measure of the difference
between the waveforms, for example.  This is cumbersome, and would
inherently depend on the inertial frame in which the waveforms are
measured.

Alternatively, we can measure the two waveforms in minimal-rotation
frames aligned with their respective radiation axes.  In that case,
Sec.~\ref{sec:EffectsOnWaveforms} showed that, again, the only freedom
to choose coordinates for the waveforms lies in the time offset
$\Delta t$ and a single phase offset rotating the system about the
radiation axis by $\Delta \Phi$.  This, of course, would \emph{not} be
the case if the frame failed to satisfy the minimal-rotation
condition.

These waveforms will also come with rotations $\Rotation{R}_{1}(t)$
and $\Rotation{R}_{2}(t)$, which describe the orientation of their
frames relative to some inertial frames.  Again, these inertial frames
need not be the same.  However, if our criteria for setting the time
and phase offsets make the waveforms equal at some fiducial time
$\tfid$, we can now directly solve for their relationship:
\begin{equation}
  \label{eq:FrameAlignment}
  \RFrame = \Rotation{R}^{-1}_{1} (\tfid)\, \Rotation{R}_{2}(\tfid
  \!+\! \Delta t)\, \e^{\Delta \Phi\, \Generator{Z}}~.
\end{equation}
This value of $\RFrame$ is then a constant, to be used at all times.
More generally, we define a new rotation operator
\begin{equation}
  \label{eq:NewRotationOperator}
  \Rotation{R}_{2}'(t) = \Rotation{R}_{2}(t+\Delta t)\, \RFrame^{-1}~.
\end{equation}
This operator is more directly comparable to the first rotation
operator, in the sense that $\Rotation{R}_{1}(\tfid) =
\Rotation{R}_{2}'(\tfid)$.

Now, waveform comparisons for non-precessing systems typically measure
differences between amplitudes and phases of the aligned waveforms.
In precessing systems decomposed in the aligned frame, it makes sense
to show the same quantities.  However, it is now also important to
examine how the frames are aligned as functions of time.  The first
interesting quantity to compare might be the angle between the two
radiation axes, $\vec{a}_{1}$ and $\vec{a}_{2}' = \Rotation{R}_{2}'\,
\hat{z}$.  A simple formula that expresses this angle is
\begin{equation}
  \label{eq:AngleBetweenRadiationAxes}
  \delta \beta \coloneqq \arccos \left[ (\Rotation{R}_{1} \hat{z})
    \cdot (\Rotation{R}_{2}' \hat{z}) \right]~.
\end{equation}
We also need to understand the relative rotation \emph{about} the
radiation axis, which is particularly important because it translates
directly into the phase of the waveform.  There is no unique way to
define this phase, whenever the two radiation axes are misaligned.  In
analogy with the above, we might use $\arccos \left[
  (\Rotation{R}_{1}\, \hat{y}) \cdot (\Rotation{R}_{2}' \hat{y})
\right]$.  However, this can be dominated by the tilt of the radiation
axes, so we attempt to remove that part of the rotation.  That is, we
define the rotation $\Rotation{R}_{1\to2'}$ that takes $\vec{a}_{1}$
into $\vec{a}_{2}'$ ``directly'' by rotating through an angle $-\delta
\beta$ about the axis $\vec{a}_{1} \times \vec{a}_{2}'$.  This can be
conveniently calculated by the axis-angle formulation or quaternions,
as described in Appendix.~\ref{sec:QuaternionForm}.  Then, the
following fulfills our needs:
\begin{equation}
  \label{eq:YawErrorAngle}
  \delta \gamma \coloneqq \arccos \left[ (\Rotation{R}_{1 \to 2'}\,
    \Rotation{R}_{1}\, \hat{y}) \cdot (\Rotation{R}_{2}' \hat{y})
  \right]~.
\end{equation}
These two angles are enough to characterize the misalignment of two
frames.

We can also use the aligned frame to construct hybrid waveforms
joining analytical and numerical waveforms in the simple manner used
for non-precessing systems~\cite{Boyle2008b}.  When hybridizing, we
use information from only the first waveform before some time $t_{1}$,
and information from only the second waveform after some time $t_{2}$,
with some transition in between.  This is typically accomplished using
a transition function $\tau(t)$ that equals 1 before $t_{1}$, 0 after
$t_{2}$, and transitions smoothly in between.  Then, for any quantity
$q$, such as the amplitude or phase of a particular mode, we define
that quantity in the hybrid as
\begin{equation}
  \label{eq:HybridQuantity}
  q_{\text{hybrid}}(t) = q_{1}(t)\, \tau(t) + q_{2}(t)\, [1 -
  \tau(t)]~.
\end{equation}
The same method does not apply trivially when applied to the rotation
operators $\Rotation{R}_{1}$ and $\Rotation{R}_{2}'$, but we can
suggest a simple method.  Various techniques have been developed by
researchers in computer graphics to allow interpolation of rotation
operators---typically with such unfortunate names as
slerp~\cite{Shoemake:1985}, nlerp, and even log-quaternion
lerp~\cite{Grassia:1998}.  Regardless of the details, any of these can
be used to define the interpolant $\Rotation{R}_{\text{interp}}(x;
t)$, which equals $\Rotation{R}_{1}(t)$ when $x=1$ and
$\Rotation{R}_{2}'(t)$ when $x=0$.  We can then define the hybrid
rotation operator as
\begin{equation}
  \label{eq:HybridRotation}
  \Rotation{R}_{\text{hybrid}}(t) = \Rotation{R}_{\text{interp}} \big(
  \tau(t); t \big)~,
\end{equation}
completing the formulation of the hybrid.

\section{Discussion}
\label{s:Discussion}

A new element has now been added to the description of a gravitational
waveform.  The complete description consists of three elements:
\begin{enumerate}
 \item specification of the inertial frame in which the waveform may
  be measured,
 \item the operator $\Rotation{R}(t)$ that rotates the inertial frame
  into the frame in which the modes of the waveform are decomposed,
  and
 \item the modes of the waveform as measured in this rotated frame.
\end{enumerate}
The new element is item 2; previously, the waveform would simply be
decomposed in the inertial frame.  References~\cite{SchmidtEtAl:2011}
and~\cite{OShaughnessyEtAl:2011} introduced criteria for choosing
$\Rotation{R}(t)$ such that the \emph{amplitudes} of the waveform
modes become simpler.  Our contribution has been the introduction of
an additional criterion which simultaneously makes the choice of
$\Rotation{R}(t)$ essentially unique and makes the \emph{phases} of
the waveform modes simpler.  Crucially, we developed the simple
formula in Eq.~\eqref{eq:MinimalRotation_Euler} for imposing our
criterion.

This change to the description of waveforms has many benefits for
precessing systems.  First, the amplitude and phase of the modes will
be more nearly approximated by smooth functions.  This means that less
storage space will be needed to describe the data from numerical
simulations, which is an increasingly important problem in
gravitational-wave modeling~\cite{AjithEtAl:2007}.  Smooth functions
are also crucial to several of the assumptions underlying
extrapolation of numerical waveforms to infinite extraction radius,
for example~\cite{Boyle-Mroue:2008}.  In general, we expect that
strong spin-spin couplings will imprint the waveform modes with some
non-smoothness, but this method does at least remove the inessential
coordinate-dependent features, as illustrated vividly in
Figs.~\ref{fig:WaveformFrequencies} and~\ref{fig:WaveformPhases}.

Second, analysis of the waveforms becomes simpler.  The machinery of
waveform manipulation for non-precessing systems is well developed.
Analytical constructions, methods for comparing waveforms to
demonstrate convergence or to measure differences from analytical
systems, and hybridization techniques are all broadly understood and
applied.  The basic approaches, however, are not designed for
precessing systems analyzed in an inertial frame; many of the
simplifying assumptions break down.  The strength of this
reformulation to include the rotation operator is that it simplifies
the waveform modes, so that we can again use techniques designed for
the non-precessing case.


\begin{acknowledgments}
  We thank Saul Teukolsky, Larry Kidder, Geoffrey Lovelace, and
  Serguei Ossokine for helpful discussions, and Richard O'Shaughnessy
  for pointing out Ref.~\cite{BuonannoEtAl:2003}. This project was
  supported in part by a grant from the Sherman Fairchild Foundation;
  by NSF Grants No.\ PHY-0969111 and No.\ PHY-1005426; and by NASA
  Grant No.\ NNX09AF96G.  H.P. gratefully acknowledges support from
  the NSERC of Canada, from the Canada Research Chairs Program, and
  from the Canadian Institute for Advanced Research.
\end{acknowledgments}

\appendix 

\section{Conventions and essential formulas}
\label{sec:Conventions}

We begin by briefly reviewing the formalism of rotations of
$\mathbb{R}^{3}$ about the origin.  Such rotations form a group,
described by orthogonal $3 \times 3$ matrices with determinant $+1$.
The group is denoted $\SO3$, and the group operation is composition of
rotations.  $\SO3$ also forms a manifold satisfying certain
consistency properties---it is a Lie
group~\cite{DuistermaatKolk:1999}.  The tangent space to that manifold
at the point corresponding to the group's identity element is called
the Lie \emph{algebra} $\so3$, consisting of all antisymmetric $3
\times 3$ matrices.  This Lie algebra is familiar as the generators of
rotations: any $\Generator{R} \in \so3$ gives rise to a rotation
$\Rotation{R} \in \SO3$ via
\begin{equation}
  \label{eq:RotationGeneratorRelationship}
  \exp : R \mapsto \Rotation{R} = \sum_{k=0}^{\infty}\, \frac{R^{k}}
  {k!}~.
\end{equation}
We also have an isomorphism between $\so3$ (with the matrix commutator
as product) and the standard 3-vectors $\mathbb{V}^{3}$ (with the
cross product).  This map is most easily represented in a standard
Cartesian frame, where the vector $\vec{v} = v^{k}\, \vec{x}_{(k)}$ is
mapped as
\begin{equation}
  \label{eq:VectorsIsomorphism}
  \sigma : v^{k} \mapsto -\epsilon^{i}_{\phantom{i}jk}\, v^{k}
  = \left(
    \begin{smallmatrix}
      0 & -v^{z} & v^{y} \\
      v^{z} & 0 & -v^{x} \\
      -v^{y} & v^{x} & 0
    \end{smallmatrix}
  \right)^{i}_{\phantom{i}j}~.
\end{equation}
In particular, we will use the isomorphism relation
\begin{equation}
  \label{eq:IsomorphismExample}
  \sigma(\vec{v} \times \vec{w}) = [\sigma(\vec{v}),
  \sigma(\vec{w})]~,
\end{equation}
for any $\vec{v}, \vec{w} \in \mathbb{V}^{3}$.  We will also use the
Cartesian basis $(\hat{x}, \hat{y}, \hat{z})$ for $\mathbb{V}^{3}$,
corresponding to the basis $(\Generator{X}, \Generator{Y},
\Generator{Z})$ for $\so3$.  We can decompose any element of $\so3$ in
this basis by translating the dot product into $\so3$:
\begin{equation}
  \label{eq:DotProduct}
  \Generator{A} \cdot \Generator{B} \coloneqq -\frac{1}{2}\,
  \sum_{i,j}\, \Generator{A}^{i}_{\phantom{i}j}\,
  \Generator{B}^{j}_{\phantom{j}i}~,
\end{equation}
which satisfies $\vec{a} \cdot \vec{b} = \sigma(\vec{a}) \cdot
\sigma(\vec{b})$.  This can be useful as in
Eq.~\eqref{eq:MinimalRotationCondition4}, for example, where the
component along $\Generator{Z}$ must be taken.

In this context, several formulas will be very useful.  For any
rotation $\Rotation{R} \in \SO3$ and vector $\vec{v} \in
\mathbb{V}^{3}$,
\begin{equation}
  \label{eq:RotationOfGenerator}
  \sigma : \Rotation{R}\, \vec{v} \mapsto \Rotation{R}\,
  \sigma(\vec{v})\, \Rotation{R}^{-1}~.
\end{equation}
For any differentiable curve $\Rotation{R}(t) \in \SO3$,
\begin{equation}
  \label{eq:DerivativeOfRotation}
  \frac{\d}{\d t} \Rotation{R}^{-1}(t) = -\Rotation{R}^{-1}(t)\,
  \dot{\Rotation{R}}(t)\, \Rotation{R}^{-1}(t)~.
\end{equation}
If $\Generator{R}(t) \in \so3$ is a curve such that $\Rotation{R}(t) =
\exp \Generator{R}(t)$, then whenever $\Generator{R}(t) \neq 0$ we can
calculate
\begin{equation}
  \label{eq:DerivativeOfRotationAsGenerator}
  \Rotation{R}^{-1}\, \dot{\Rotation{R}} = \dot{\Generator{R}} -
  \frac{1 - \cos r}{r^{2}}\, [\Generator{R}, \dot{\Generator{R}}] +
  \frac{r - \sin r}{r^{3}}\, \big[\Generator{R}, [\Generator{R},
  \dot{\Generator{R}}] \big]~,
\end{equation}
where $r$ is the magnitude of the nonzero eigenvalues of
$\Generator{R}$, which also equals the vector norm $\abs{\vec{r}} =
\abs{\sigma^{-1}( \Generator{R} )}$.  For any $t$ such that
$\Generator{R}(t)=0$, only the first term remains.  Finally, for
$\Generator{A}, \Generator{B} \in \so3$ with $\Generator{A} \neq 0$,
\begin{equation}
  \label{eq:GeneratorConjugation}
  \e^{A}\, B\, \e^{-A} = B + \frac{1-\cos a}{a^{2}} \big[A, [A,B]
  \big] + \frac{\sin a}{a}\, [A, B]~,
\end{equation}
where $a$ is similarly the magnitude of the nonzero eigenvalues of
$\Generator{A}$.  Obviously, when $\Generator{A}=0$, only the first
term remains.

Euler angles form one convenient set of coordinates for $\SO3$.  We
define the Euler angles using the $z$-$y'$-$z''$
convention,\footnote{Note that the $z$-$x'$-$z''$ convention is more
  standard in classical mechanics---as in
  Ref.~\cite{GoldsteinEtAl:2001}, for example.  The $z$-$y'$-$z''$
  convention, however, is more standard when using spherical
  harmonics, as we do here.} where the first rotation is through an
angle $\alpha$ about the $z$ axis, the second through $\beta$ about
the (new) $y'$ axis, and the third through $\gamma$ about the (new)
$z''$ axis.  Note that this is equivalent to rotations in the opposite
order about the \emph{fixed} set of axes $z$-$y$-$z$---which is an
especially useful form for calculations.  In particular, we can
express the rotation operator as
\begin{equation}
  \label{eq:RotationOperator}
  \Rotation{R}(\alpha, \beta, \gamma) = \e^{\alpha\, \Generator{Z}}\,
  \e^{\beta\, \Generator{Y}}\, \e^{\gamma\, \Generator{Z}}~.
\end{equation}
Note that, in this convention, the singularities of the Euler angle
coordinates occur when $\beta=0$ or $\beta = \pi$.

The Wigner matrices $\mathcal{D}^{(\ell)}$ form a representation of
the rotation group.  That is, we know that a composition of rotations
in $\SO3$ given by
\begin{equation}
  \label{eq:CompositionOfRotations}
  \Rotation{R}(\alpha, \beta, \gamma) = \Rotation{R}(\alpha', \beta',
  \gamma') \, \Rotation{R}(\alpha'', \beta'', \gamma'')
\end{equation}
implies the relation
\begin{multline}
  \label{eq:CompositionOfWignerMatrices}
  \mathcal{D}^{(\ell)}_{m',m}(\alpha, \beta, \gamma) = \\ \sum_{m''}\,
  \mathcal{D}^{(\ell)}_{m',m''}(\alpha', \beta', \gamma') \,
  \mathcal{D}^{(\ell)}_{m'',m}(\alpha'', \beta'', \gamma'')~.
\end{multline}
We can find an explicit formula for the $\mathcal{D}^{(\ell)}$ as in
Ref.~\cite{Wigner:1959} which, in our conventions, gives
\begin{multline}
  \label{eq:WignerDMatrices}
  \mathcal{D}^{(\ell)}_{m',m}(\alpha, \beta, \gamma) = (-1)^{\ell +
    m}\, \sqrt{ \frac{(\ell+m)!(\ell-m)!} {(\ell+m')!(\ell-m')!} } \\
  \times \e^{\i(m\, \alpha + m'\, \gamma)}\, \sum_{\rho}\,
  (-1)^{\rho}\, \binom{\ell + m'} {\rho}\,
  \binom{\ell-m'} {\rho-m-m'} \\
  \times \sin\left( \frac{\beta}{2} \right)^{2\ell - 2\rho + m + m'}\,
  \cos\left( \frac{\beta}{2} \right)^{2\rho - m - m'}~.
\end{multline}

Goldberg \etal~\cite{GoldbergEtAl:1967} showed that the spin-weighted
spherical harmonics~\cite{NewmanPenrose:1966} are simply special cases
of this expression.  Adopting conventions to agree with
Ref.~\cite{AjithEtAl:2007}, we have
\begin{equation}
  \label{eq:SWSHs}
  \sYlm{\ell,m}(\vartheta, \varphi) = (-1)^{m}\, \sqrt{ \frac{2\, \ell
      + 1} {4\, \pi} }\, \mathcal{D}^{(\ell)}_{m,-s}(0, \vartheta,
  \varphi)~.
\end{equation}
Combining Eqs.~\eqref{eq:CompositionOfWignerMatrices}
and~\eqref{eq:SWSHs}, we can immediately obtain the transformation law
for the spherical harmonics under rotation of the
coordinates~\cite{BoyleThesis}:
\begin{equation}
  \label{eq:SWSHRotation}
  \sYlm{\ell, m'}(\vartheta', \varphi') = \sum_{m}
  \mathcal{D}^{(\ell)}_{m',m}(\alpha, \beta, \gamma)\,
  \sYlm{\ell,m}(\vartheta, \varphi)~,
\end{equation}
for angles satisfying
\begin{equation}
  \label{eq:AngleRelationship}
  \Rotation{R}(0, \vartheta', \varphi') = \Rotation{R}(\alpha, \beta,
  \gamma)\, \Rotation{R}(0, \vartheta, \varphi)~.
\end{equation}
Note that the rotation $\Rotation{R}(\alpha, \beta, \gamma)$ is
uniquely fixed by this condition in a way that it would not be by
simply requiring the point defined by $(\vartheta', \varphi')$ to
rotate into the point defined by $(\vartheta, \varphi)$.  The fact
that such a condition on the coordinates would not uniquely define
$\Rotation{R}(\alpha, \beta, \gamma)$ is the central problem addressed
by this paper.

We decompose a function $q$ of spin weight $s$ as
\begin{equation}
  \label{eq:SWSHDecomposition}
  \wavefield(\vartheta, \varphi) = \sum_{\ell, m}\,
  \wavefield^{\ell,m}\, \sYlm{\ell,m}(\vartheta, \varphi)~,
\end{equation}
We then define the rotated function $\Rotated{\wavefield}$ satisfying
$\Rotated{\wavefield}(\vartheta', \varphi') = \wavefield(\vartheta,
\varphi)$, where the angles are again related by
Eq.~\eqref{eq:AngleRelationship}.  We can decompose
$\Rotated{\wavefield}$ into spin-weighted spherical harmonics in
$(\vartheta', \varphi')$, and use Eq.~\eqref{eq:SWSHRotation} to find
the relations between the components:
\begin{subequations}
  \label{eq:FieldTransformations}
  \begin{equation}
    \label{eq:FieldTransformationInv}
    \wavefield^{\ell, m} = \sum_{m'}\, \Rotated{\wavefield}^{\ell,
      m'}\, \mathcal{D}^{(\ell)}_{m',m}(\alpha, \beta, \gamma)~,
  \end{equation}
  or equivalently
  \begin{equation}
    \label{eq:FieldTransformation}
    \Rotated{\wavefield}^{\ell, m} = \sum_{m'}\, \wavefield^{\ell,
      m'}\, \mathcal{D}^{(\ell)}_{m',m}(-\gamma, -\beta, -\alpha)~.
  \end{equation}
\end{subequations}
These relations are special cases of more general transformations
derived by Gualtieri \etal~\cite{GualtieriEtAl:2008}.

\section{Main results in quaternion form}
\label{sec:QuaternionForm}

Another useful parameterization of the rotation group is given by the
set of unit quaternions, which are useful as an efficient numerical
method for representing rotations.  Here, we repeat the main results
of this paper (the minimal-rotation condition and a method for
imposing it) in the form of quaternions.  We also express the Wigner
matrices $\mathcal{D}^{(\ell)}$ as functions of a quaternion, instead
of Euler angles.

Quaternions may be thought of in many ways, but the form we find
convenient here is that of a scalar plus a vector.  The notation we
use will be $Q = q_{0} + \vec{q} = (q_{0}, q_{1}, q_{2}, q_{3})$,
where $q_{0}$ is the scalar part, $\vec{q}$ is the vector part, and
$(q_{1}, q_{2}, q_{3})$ are the Cartesian components of the vector.
The conjugate of the quaternion is $\co{Q} = q_{0} - \vec{q}$.  The
distinctive feature of quaternions is their unusual multiplication
rule:
\begin{equation}
  \label{eq:QuaternionMultiplication}
  P\, Q = (p_{0}\, q_{0} - \vec{p} \cdot \vec{q}) + ( p_{0}\, \vec{q}
  + q_{0}\, \vec{p} + \vec{p} \times \vec{q})~.
\end{equation}
Note that this product is associative (unlike the vector cross
product) but is neither commutative nor anti-commutative for general
quaternions.  The unit quaternions satisfy $Q\, \co{Q} = 1$, and can
be thought of as points on the unit 3-sphere.  The set of all unit
quaternions forms a group locally isomorphic to $\SO3$, where rotation
of a vector $\vec{v}$ by the unit quaternion $Q$ can be expressed by
\begin{equation}
  \label{eq:QuaternionRotation}
  \vec{v}' = Q\, \vec{v}\, \co{Q}~.
\end{equation}
In fact, the quaternions provide a double cover of $\SO3$, as can be
seen above by the fact that $Q$ and $-Q$ produce the same rotation,
but this does not cause any practical problems.  The relationship with
the axis-angle formulation of rotations is particularly clean: for a
rotation through an angle $\delta$ about an axis $\hat{w}$, the
quaternion is given by $Q = \cos(\delta/2) + \sin(\delta/2)\,
\hat{w}$.  In any case, we may think of a general rotation
$\Rotation{R}$ as being precisely equivalent to some unit quaternion.
For further details, we refer to standard texts (e.g.,
Ref.~\cite{DoranLasenby:2003}).

A calculation very similar to the derivation of
Eq.~\eqref{eq:FrameRotationGenerator} shows us that the instantaneous
rotation vector associated to the unit quaternion $R$ is given by
\begin{equation}
  \label{eq:FrameRotationGenerator_Quaternion}
  \FrameRot = 2\, \dot{R}\, \co{R}~.
\end{equation}
If $R$ rotates the $z$ axis into the radiation axis, $\vec{a} = R\,
\hat{z}\, \co{R}$, then the minimal-rotation condition given by
Eq.~\eqref{eq:MinimalRotationCondition} can be rewritten as
\begin{equation}
  \label{eq:MinimalRotationCondition_Quaternions}
  ( \dot{R}\, \hat{z}\, \co{R} )_{0} = 0~,
\end{equation}
where the subscript on the left-hand side takes the scalar part of the
expression.  This equation is equivalent to the condition on the
rotation operator itself given in
Eq.~\eqref{eq:MinimalRotationCondition4}.

As before, we may impose this condition by applying an initial
rotation about the $z$ axis.  Thus, if $\YawfulRotQ$ is any quaternion
rotation that takes the $z$ axis into the radiation axis, we define
the rotation $R = \YawfulRotQ\, \GammaRotQ$, where $\GammaRotQ$ is a
rotation through an angle $\gamma$ about the $z$ axis.  Then $R$
satisfies the minimal-rotation condition if
\begin{equation}
  \label{eq:MinimalRotation_Quaternion}
  \dot{\gamma} = 2 ( \YawfulRotDotQ\, \hat{z}\, \YawfulRotCoQ)_{0}~.
\end{equation}
Again, given the result $\YawfulRotQ$ of some axis-alignment method,
we can evaluate the right-hand side, integrate, and construct the
total rotation.

Finally, if the result is to be used to rotate modes of a waveform, we
need to express the Wigner matrices in quaternion form.  This is done
most simply by relating the Euler angles to various components of the
quaternion, and substituting the results in
Eq.~\eqref{eq:WignerDMatrices}.  We find
\begin{multline}
  \label{eq:WignerDMatrices_Quaternions}
  \mathcal{D}^{(\ell)}_{m',m}(R) = (-2)^{m-\ell}\, \sqrt{
    \frac{(\ell+m)!(\ell-m)!} {(\ell+m')!(\ell-m')!} } \\
  \times \left[ (R\hat{z})_{3} - \i\, (R\hat{z})_{0} \right]^{m+m'}\,
  \left[ (R\hat{z})_{1} + \i\, (R\hat{z})_{2} \right]^{m-m'} \\
  \times \sum_{\rho}\, (-1)^{\rho}\, \binom{\ell + m'} {\rho}\,
  \binom{\ell-m'} {\rho-m-m'} \\
  \times \left[ 1 - (R \hat{z} \bar{R})_{3} \right]^{\ell + m' -
    \rho}\, \left[ 1 + (R \hat{z} \bar{R})_{3} \right]^{\rho - m -
    m'}~.
\end{multline}
This expression can be used in Eq.~\eqref{eq:FieldTransformationInv};
for Eq.~\eqref{eq:FieldTransformation}, $\co{R} = R^{-1}$ should
replace $R$ in this expression.

\let\c\Originalcdefinition %
\let\d\Originalddefinition %
\let\i\Originalidefinition %
\bibliography{References}

\begin{thebibliography}{37}%
\makeatletter
\providecommand \@ifxundefined [1]{%
 \@ifx{#1\undefined}
}%
\providecommand \@ifnum [1]{%
 \ifnum #1\expandafter \@firstoftwo
 \else \expandafter \@secondoftwo
 \fi
}%
\providecommand \@ifx [1]{%
 \ifx #1\expandafter \@firstoftwo
 \else \expandafter \@secondoftwo
 \fi
}%
\providecommand \natexlab [1]{#1}%
\providecommand \enquote  [1]{``#1''}%
\providecommand \bibnamefont  [1]{#1}%
\providecommand \bibfnamefont [1]{#1}%
\providecommand \citenamefont [1]{#1}%
\providecommand \href@noop [0]{\@secondoftwo}%
\providecommand \href [0]{\begingroup \@sanitize@url \@href}%
\providecommand \@href[1]{\@@startlink{#1}\@@href}%
\providecommand \@@href[1]{\endgroup#1\@@endlink}%
\providecommand \@sanitize@url [0]{\catcode `\\12\catcode `\$12\catcode
  `\&12\catcode `\#12\catcode `\^12\catcode `\_12\catcode `\%12\relax}%
\providecommand \@@startlink[1]{}%
\providecommand \@@endlink[0]{}%
\providecommand \url  [0]{\begingroup\@sanitize@url \@url }%
\providecommand \@url [1]{\endgroup\@href {#1}{\urlprefix }}%
\providecommand \urlprefix  [0]{URL }%
\providecommand \Eprint [0]{\href }%
\providecommand \doibase [0]{http://dx.doi.org/}%
\providecommand \selectlanguage [0]{\@gobble}%
\providecommand \bibinfo  [0]{\@secondoftwo}%
\providecommand \bibfield  [0]{\@secondoftwo}%
\providecommand \translation [1]{[#1]}%
\providecommand \BibitemOpen [0]{}%
\providecommand \bibitemStop [0]{}%
\providecommand \bibitemNoStop [0]{.\EOS\space}%
\providecommand \EOS [0]{\spacefactor3000\relax}%
\providecommand \BibitemShut  [1]{\csname bibitem#1\endcsname}%
\let\auto@bib@innerbib\@empty
\bibitem [{\citenamefont {Finn}(1992)}]{Finn:1992}%
  \BibitemOpen
  \bibfield  {author} {\bibinfo {author} {\bibfnamefont {L.~S.}\ \bibnamefont
  {Finn}},\ }\href {\doibase 10.1103/PhysRevD.46.5236} {\bibfield  {journal}
  {\bibinfo  {journal} {Phys. Rev. D}\ }\textbf {\bibinfo {volume} {46}},\
  \bibinfo {pages} {5236} (\bibinfo {year} {1992})}\BibitemShut {NoStop}%
\bibitem [{\citenamefont {Finn}\ and\ \citenamefont
  {Chernoff}(1993)}]{FinnChernoff:1993}%
  \BibitemOpen
  \bibfield  {author} {\bibinfo {author} {\bibfnamefont {L.~S.}\ \bibnamefont
  {Finn}}\ and\ \bibinfo {author} {\bibfnamefont {D.~F.}\ \bibnamefont
  {Chernoff}},\ }\href {\doibase 10.1103/PhysRevD.47.2198} {\bibfield
  {journal} {\bibinfo  {journal} {Phys. Rev. D}\ }\textbf {\bibinfo {volume}
  {47}},\ \bibinfo {pages} {2198} (\bibinfo {year} {1993})}\BibitemShut
  {NoStop}%
\bibitem [{\citenamefont {Barish}\ and\ \citenamefont
  {Weiss}(1999)}]{Barish:1999}%
  \BibitemOpen
  \bibfield  {author} {\bibinfo {author} {\bibfnamefont {B.~C.}\ \bibnamefont
  {Barish}}\ and\ \bibinfo {author} {\bibfnamefont {R.}~\bibnamefont {Weiss}},\
  }\href {\doibase 10.1063/1.882861} {\bibfield  {journal} {\bibinfo  {journal}
  {Phys. Today}\ }\textbf {\bibinfo {volume} {52}},\ \bibinfo {pages} {44}
  (\bibinfo {year} {1999})}\BibitemShut {NoStop}%
\bibitem [{\citenamefont {Sigg}\ and\ \citenamefont {the {LIGO}
  Scientific~Collaboration}(2008)}]{Sigg:2008}%
  \BibitemOpen
  \bibfield  {author} {\bibinfo {author} {\bibfnamefont {D.}~\bibnamefont
  {Sigg}}\ and\ \bibinfo {author} {\bibnamefont {the {LIGO}
  Scientific~Collaboration}},\ }\href {\doibase 10.1088/0264-9381/25/11/114041}
  {\bibfield  {journal} {\bibinfo  {journal} {Class. Quant. Grav.}\ }\textbf
  {\bibinfo {volume} {25}},\ \bibinfo {pages} {114041} (\bibinfo {year}
  {2008})}\BibitemShut {NoStop}%
\bibitem [{\citenamefont {Acernese}\ \emph {et~al.}(2008)\citenamefont
  {Acernese} \emph {et~al.}}]{Acernese:2008}%
  \BibitemOpen
  \bibfield  {author} {\bibinfo {author} {\bibfnamefont {F.}~\bibnamefont
  {Acernese}} \emph {et~al.},\ }\href {\doibase 10.1088/0264-9381/25/18/184001}
  {\bibfield  {journal} {\bibinfo  {journal} {Class. Quant. Grav.}\ }\textbf
  {\bibinfo {volume} {25}},\ \bibinfo {pages} {184001} (\bibinfo {year}
  {2008})}\BibitemShut {NoStop}%
\bibitem [{\citenamefont {Kuroda}\ and\ \citenamefont {the
  {LCGT}~Collaboration}(2010)}]{Kuroda:2010}%
  \BibitemOpen
  \bibfield  {author} {\bibinfo {author} {\bibfnamefont {K.}~\bibnamefont
  {Kuroda}}\ and\ \bibinfo {author} {\bibnamefont {the {LCGT}~Collaboration}},\
  }\href {\doibase 10.1088/0264-9381/27/8/084004} {\bibfield  {journal}
  {\bibinfo  {journal} {Class. Quant. Grav.}\ }\textbf {\bibinfo {volume}
  {27}},\ \bibinfo {pages} {084004} (\bibinfo {year} {2010})}\BibitemShut
  {NoStop}%
\bibitem [{\citenamefont {{Prince}}\ \emph {et~al.}(2006)\citenamefont
  {{Prince}}, \citenamefont {{Binetruy}}, \citenamefont {{Centrella}},
  \citenamefont {{Finn}}, \citenamefont {{Hogan}}, \citenamefont {{Nelemans}},
  \citenamefont {{Phinney}}, \citenamefont {{Schutz}},\ and\ \citenamefont
  {{LISA International Science Team}}}]{Lisa}%
  \BibitemOpen
  \bibfield  {author} {\bibinfo {author} {\bibfnamefont {T.~A.}\ \bibnamefont
  {{Prince}}}, \bibinfo {author} {\bibfnamefont {P.}~\bibnamefont
  {{Binetruy}}}, \bibinfo {author} {\bibfnamefont {J.}~\bibnamefont
  {{Centrella}}}, \bibinfo {author} {\bibfnamefont {L.~S.}\ \bibnamefont
  {{Finn}}}, \bibinfo {author} {\bibfnamefont {C.}~\bibnamefont {{Hogan}}},
  \bibinfo {author} {\bibfnamefont {G.}~\bibnamefont {{Nelemans}}}, \bibinfo
  {author} {\bibfnamefont {E.~S.}\ \bibnamefont {{Phinney}}}, \bibinfo {author}
  {\bibfnamefont {B.}~\bibnamefont {{Schutz}}}, \ and\ \bibinfo {author}
  {\bibnamefont {{LISA International Science Team}}},\ }in\ \href
  {http://adsabs.harvard.edu/abs/2006AAS...209.7401P} {\emph {\bibinfo
  {booktitle} {American Astronomical Society Meeting Abstracts}}},\ \bibinfo
  {series} {Bulletin of the American Astronomical Society}, Vol.~\bibinfo
  {volume} {38}\ (\bibinfo {year} {2006})\ p.\ \bibinfo {pages}
  {990}\BibitemShut {NoStop}%
\bibitem [{\citenamefont {Bender}\ \emph {et~al.}(1998)\citenamefont {Bender}
  \emph {et~al.}}]{Lisa98}%
  \BibitemOpen
  \bibfield  {author} {\bibinfo {author} {\bibfnamefont {P.~L.}\ \bibnamefont
  {Bender}} \emph {et~al.},\ }\href
  {http://pubman.mpdl.mpg.de/pubman/faces/viewItemFullPage.jsp?itemId=escidoc:52082}
  {\emph {\bibinfo {title} {{LISA.} {L}aser {I}nterferometer {S}pace {A}ntenna
  for the detection and observation of gravitational waves}}},\ \bibinfo {type}
  {Tech. Rep.}\ (\bibinfo  {institution} {{Max-Planck-Institut} f{\"{u}}r
  Quantenoptik},\ \bibinfo {address} {M{\"{u}}nchen, Germany},\ \bibinfo {year}
  {1998})\BibitemShut {NoStop}%
\bibitem [{\citenamefont {Jennrich}(2009)}]{Jennrich:2009}%
  \BibitemOpen
  \bibfield  {author} {\bibinfo {author} {\bibfnamefont {O.}~\bibnamefont
  {Jennrich}},\ }\href {\doibase 10.1088/0264-9381/26/15/153001} {\bibfield
  {journal} {\bibinfo  {journal} {Class. Quant. Grav.}\ }\textbf {\bibinfo
  {volume} {26}},\ \bibinfo {pages} {153001} (\bibinfo {year}
  {2009})}\BibitemShut {NoStop}%
\bibitem [{\citenamefont {Centrella}\ \emph {et~al.}(2010)\citenamefont
  {Centrella}, \citenamefont {Baker}, \citenamefont {Kelly},\ and\
  \citenamefont {van Meter}}]{Centrella:2010}%
  \BibitemOpen
  \bibfield  {author} {\bibinfo {author} {\bibfnamefont {J.}~\bibnamefont
  {Centrella}}, \bibinfo {author} {\bibfnamefont {J.~G.}\ \bibnamefont
  {Baker}}, \bibinfo {author} {\bibfnamefont {B.~J.}\ \bibnamefont {Kelly}}, \
  and\ \bibinfo {author} {\bibfnamefont {J.~R.}\ \bibnamefont {van Meter}},\
  }\href {\doibase 10.1103/RevModPhys.82.3069} {\bibfield  {journal} {\bibinfo
  {journal} {Rev. Mod. Phys.}\ }\textbf {\bibinfo {volume} {82}},\ \bibinfo
  {pages} {3069} (\bibinfo {year} {2010})}\BibitemShut {NoStop}%
\bibitem [{\citenamefont {{McWilliams}}(2011)}]{McWilliams:2010iq}%
  \BibitemOpen
  \bibfield  {author} {\bibinfo {author} {\bibfnamefont {S.~T.}\ \bibnamefont
  {{McWilliams}}},\ }\href {\doibase 10.1088/0264-9381/28/13/134001} {\bibfield
   {journal} {\bibinfo  {journal} {Class. Quant. Grav.}\ }\textbf {\bibinfo
  {volume} {28}},\ \bibinfo {pages} {134001} (\bibinfo {year}
  {2011})}\BibitemShut {NoStop}%
\bibitem [{\citenamefont {Lousto}\ and\ \citenamefont
  {Zlochower}(2011)}]{LoustoZlochower2010}%
  \BibitemOpen
  \bibfield  {author} {\bibinfo {author} {\bibfnamefont {C.~O.}\ \bibnamefont
  {Lousto}}\ and\ \bibinfo {author} {\bibfnamefont {Y.}~\bibnamefont
  {Zlochower}},\ }\href
  {http://link.aps.org/doi/10.1103/PhysRevLett.106.041101} {\bibfield
  {journal} {\bibinfo  {journal} {Phys. Rev. Lett.}\ }\textbf {\bibinfo
  {volume} {106}},\ \bibinfo {pages} {041101} (\bibinfo {year}
  {2011})}\BibitemShut {NoStop}%
\bibitem [{\citenamefont {Lovelace}\ \emph
  {et~al.}(2011{\natexlab{a}})\citenamefont {Lovelace}, \citenamefont
  {Scheel},\ and\ \citenamefont {Szil{\'{a}}gyi}}]{Lovelace2010}%
  \BibitemOpen
  \bibfield  {author} {\bibinfo {author} {\bibfnamefont {G.}~\bibnamefont
  {Lovelace}}, \bibinfo {author} {\bibfnamefont {M.~A.}\ \bibnamefont
  {Scheel}}, \ and\ \bibinfo {author} {\bibfnamefont {B.}~\bibnamefont
  {Szil{\'{a}}gyi}},\ }\href {\doibase 10.1103/PhysRevD.83.024010} {\bibfield
  {journal} {\bibinfo  {journal} {Phys. Rev. D}\ }\textbf {\bibinfo {volume}
  {83}},\ \bibinfo {pages} {024010} (\bibinfo {year}
  {2011}{\natexlab{a}})}\BibitemShut {NoStop}%
\bibitem [{\citenamefont {Lovelace}\ \emph
  {et~al.}(2011{\natexlab{b}})\citenamefont {Lovelace}, \citenamefont {Boyle},
  \citenamefont {Scheel},\ and\ \citenamefont
  {Szil{\'{a}}gyi}}]{Lovelace:2011}%
  \BibitemOpen
  \bibfield  {author} {\bibinfo {author} {\bibfnamefont {G.}~\bibnamefont
  {Lovelace}}, \bibinfo {author} {\bibfnamefont {M.}~\bibnamefont {Boyle}},
  \bibinfo {author} {\bibfnamefont {M.~A.}\ \bibnamefont {Scheel}}, \ and\
  \bibinfo {author} {\bibfnamefont {B.}~\bibnamefont {Szil{\'{a}}gyi}},\
  }\href@noop {} {\enquote {\bibinfo {title} {Accurate gravitational waveforms
  for binary-black-hole mergers with nearly extremal spins},}\ } (\bibinfo
  {year} {2011}{\natexlab{b}}),\ \Eprint {http://arxiv.org/abs/1110.2229}
  {arXiv:1110.2229 [gr-qc]} \BibitemShut {NoStop}%
\bibitem [{\citenamefont {Apostolatos}\ \emph {et~al.}(1994)\citenamefont
  {Apostolatos}, \citenamefont {Cutler}, \citenamefont {Sussman},\ and\
  \citenamefont {Thorne}}]{ApostolatosEtAl:1994}%
  \BibitemOpen
  \bibfield  {author} {\bibinfo {author} {\bibfnamefont {T.~A.}\ \bibnamefont
  {Apostolatos}}, \bibinfo {author} {\bibfnamefont {C.}~\bibnamefont {Cutler}},
  \bibinfo {author} {\bibfnamefont {G.~J.}\ \bibnamefont {Sussman}}, \ and\
  \bibinfo {author} {\bibfnamefont {K.~S.}\ \bibnamefont {Thorne}},\ }\href
  {\doibase 10.1103/PhysRevD.49.6274} {\bibfield  {journal} {\bibinfo
  {journal} {Phys. Rev. D}\ }\textbf {\bibinfo {volume} {49}},\ \bibinfo
  {pages} {6274} (\bibinfo {year} {1994})}\BibitemShut {NoStop}%
\bibitem [{\citenamefont {Kidder}(1995)}]{Kidder:1995}%
  \BibitemOpen
  \bibfield  {author} {\bibinfo {author} {\bibfnamefont {L.~E.}\ \bibnamefont
  {Kidder}},\ }\href {\doibase 10.1103/PhysRevD.52.821} {\bibfield  {journal}
  {\bibinfo  {journal} {Phys. Rev. D}\ }\textbf {\bibinfo {volume} {52}},\
  \bibinfo {pages} {821} (\bibinfo {year} {1995})}\BibitemShut {NoStop}%
\bibitem [{\citenamefont {Gualtieri}\ \emph {et~al.}(2008)\citenamefont
  {Gualtieri}, \citenamefont {Berti}, \citenamefont {Cardoso},\ and\
  \citenamefont {Sperhake}}]{GualtieriEtAl:2008}%
  \BibitemOpen
  \bibfield  {author} {\bibinfo {author} {\bibfnamefont {L.}~\bibnamefont
  {Gualtieri}}, \bibinfo {author} {\bibfnamefont {E.}~\bibnamefont {Berti}},
  \bibinfo {author} {\bibfnamefont {V.}~\bibnamefont {Cardoso}}, \ and\
  \bibinfo {author} {\bibfnamefont {U.}~\bibnamefont {Sperhake}},\ }\href
  {\doibase 10.1103/PhysRevD.78.044024} {\bibfield  {journal} {\bibinfo
  {journal} {Phys. Rev. D}\ }\textbf {\bibinfo {volume} {78}},\ \bibinfo
  {pages} {044024} (\bibinfo {year} {2008})}\BibitemShut {NoStop}%
\bibitem [{\citenamefont {Schmidt}\ \emph {et~al.}(2011)\citenamefont
  {Schmidt}, \citenamefont {Hannam}, \citenamefont {Husa},\ and\ \citenamefont
  {Ajith}}]{SchmidtEtAl:2011}%
  \BibitemOpen
  \bibfield  {author} {\bibinfo {author} {\bibfnamefont {P.}~\bibnamefont
  {Schmidt}}, \bibinfo {author} {\bibfnamefont {M.}~\bibnamefont {Hannam}},
  \bibinfo {author} {\bibfnamefont {S.}~\bibnamefont {Husa}}, \ and\ \bibinfo
  {author} {\bibfnamefont {P.}~\bibnamefont {Ajith}},\ }\href {\doibase
  10.1103/PhysRevD.84.024046} {\bibfield  {journal} {\bibinfo  {journal} {Phys.
  Rev. D}\ }\textbf {\bibinfo {volume} {84}},\ \bibinfo {pages} {024046}
  (\bibinfo {year} {2011})}\BibitemShut {NoStop}%
\bibitem [{\citenamefont {{O'Shaughnessy}}\ \emph {et~al.}(2011)\citenamefont
  {{O'Shaughnessy}}, \citenamefont {Vaishnav}, \citenamefont {Healy},
  \citenamefont {Meeks},\ and\ \citenamefont
  {Shoemaker}}]{OShaughnessyEtAl:2011}%
  \BibitemOpen
  \bibfield  {author} {\bibinfo {author} {\bibfnamefont {R.}~\bibnamefont
  {{O'Shaughnessy}}}, \bibinfo {author} {\bibfnamefont {B.}~\bibnamefont
  {Vaishnav}}, \bibinfo {author} {\bibfnamefont {J.}~\bibnamefont {Healy}},
  \bibinfo {author} {\bibfnamefont {Z.}~\bibnamefont {Meeks}}, \ and\ \bibinfo
  {author} {\bibfnamefont {D.}~\bibnamefont {Shoemaker}},\ }\href@noop {}
  {\enquote {\bibinfo {title} {Efficient asymptotic frame selection for binary
  black hole spacetimes using asymptotic radiation},}\ } (\bibinfo {year}
  {2011}),\ \Eprint {http://arxiv.org/abs/1109.5224} {arXiv:1109.5224 [gr-qc]}
  \BibitemShut {NoStop}%
\bibitem [{\citenamefont {Buonanno}\ \emph {et~al.}(2003)\citenamefont
  {Buonanno}, \citenamefont {Chen},\ and\ \citenamefont
  {Vallisneri}}]{BuonannoEtAl:2003}%
  \BibitemOpen
  \bibfield  {author} {\bibinfo {author} {\bibfnamefont {A.}~\bibnamefont
  {Buonanno}}, \bibinfo {author} {\bibfnamefont {Y.}~\bibnamefont {Chen}}, \
  and\ \bibinfo {author} {\bibfnamefont {M.}~\bibnamefont {Vallisneri}},\
  }\href {\doibase 10.1103/PhysRevD.67.104025} {\bibfield  {journal} {\bibinfo
  {journal} {Phys. Rev. D}\ }\textbf {\bibinfo {volume} {67}},\ \bibinfo
  {pages} {104025} (\bibinfo {year} {2003})}\BibitemShut {NoStop}%
\bibitem [{\citenamefont {Goldberg}\ \emph {et~al.}(1967)\citenamefont
  {Goldberg}, \citenamefont {Macfarlane}, \citenamefont {Newman}, \citenamefont
  {Rohrlich},\ and\ \citenamefont {Sudarshan}}]{GoldbergEtAl:1967}%
  \BibitemOpen
  \bibfield  {author} {\bibinfo {author} {\bibfnamefont {J.~N.}\ \bibnamefont
  {Goldberg}}, \bibinfo {author} {\bibfnamefont {A.~J.}\ \bibnamefont
  {Macfarlane}}, \bibinfo {author} {\bibfnamefont {E.~T.}\ \bibnamefont
  {Newman}}, \bibinfo {author} {\bibfnamefont {F.}~\bibnamefont {Rohrlich}}, \
  and\ \bibinfo {author} {\bibfnamefont {E.~C.~G.}\ \bibnamefont {Sudarshan}},\
  }\href {\doibase 10.1063/1.1705135} {\bibfield  {journal} {\bibinfo
  {journal} {J. Math. Phys.}\ }\textbf {\bibinfo {volume} {8}},\ \bibinfo
  {pages} {2155} (\bibinfo {year} {1967})}\BibitemShut {NoStop}%
\bibitem [{\citenamefont {Duistermaat}\ and\ \citenamefont
  {Kolk}(1999)}]{DuistermaatKolk:1999}%
  \BibitemOpen
  \bibfield  {author} {\bibinfo {author} {\bibfnamefont {J.~J.}\ \bibnamefont
  {Duistermaat}}\ and\ \bibinfo {author} {\bibfnamefont {J.~A.~C.}\
  \bibnamefont {Kolk}},\ }\href@noop {} {\emph {\bibinfo {title} {Lie
  Groups}}}\ (\bibinfo  {publisher} {Springer-Verlag},\ \bibinfo {address} {New
  York, NY},\ \bibinfo {year} {1999})\BibitemShut {NoStop}%
\bibitem [{\citenamefont {Arun}\ \emph {et~al.}(2009)\citenamefont {Arun},
  \citenamefont {Buonanno}, \citenamefont {Faye},\ and\ \citenamefont
  {Ochsner}}]{ArunEtAl:2009}%
  \BibitemOpen
  \bibfield  {author} {\bibinfo {author} {\bibfnamefont {K.~G.}\ \bibnamefont
  {Arun}}, \bibinfo {author} {\bibfnamefont {A.}~\bibnamefont {Buonanno}},
  \bibinfo {author} {\bibfnamefont {G.}~\bibnamefont {Faye}}, \ and\ \bibinfo
  {author} {\bibfnamefont {E.}~\bibnamefont {Ochsner}},\ }\href {\doibase
  10.1103/PhysRevD.79.104023} {\bibfield  {journal} {\bibinfo  {journal} {Phys.
  Rev. D}\ }\textbf {\bibinfo {volume} {79}},\ \bibinfo {pages} {104023}
  (\bibinfo {year} {2009})}\BibitemShut {NoStop}%
\bibitem [{\citenamefont {Campanelli}\ \emph {et~al.}(2009)\citenamefont
  {Campanelli}, \citenamefont {Lousto}, \citenamefont {Nakano},\ and\
  \citenamefont {Zlochower}}]{CampanelliEtAl:2009}%
  \BibitemOpen
  \bibfield  {author} {\bibinfo {author} {\bibfnamefont {M.}~\bibnamefont
  {Campanelli}}, \bibinfo {author} {\bibfnamefont {C.~O.}\ \bibnamefont
  {Lousto}}, \bibinfo {author} {\bibfnamefont {H.}~\bibnamefont {Nakano}}, \
  and\ \bibinfo {author} {\bibfnamefont {Y.}~\bibnamefont {Zlochower}},\ }\href
  {\doibase 10.1103/PhysRevD.79.084010} {\bibfield  {journal} {\bibinfo
  {journal} {Phys. Rev. D}\ }\textbf {\bibinfo {volume} {79}},\ \bibinfo
  {pages} {084010} (\bibinfo {year} {2009})}\BibitemShut {NoStop}%
\bibitem [{\citenamefont {Blanchet}\ \emph {et~al.}(2008)\citenamefont
  {Blanchet}, \citenamefont {Faye}, \citenamefont {Iyer},\ and\ \citenamefont
  {Sinha}}]{BlanchetEtAl:2008}%
  \BibitemOpen
  \bibfield  {author} {\bibinfo {author} {\bibfnamefont {L.}~\bibnamefont
  {Blanchet}}, \bibinfo {author} {\bibfnamefont {G.}~\bibnamefont {Faye}},
  \bibinfo {author} {\bibfnamefont {B.~R.}\ \bibnamefont {Iyer}}, \ and\
  \bibinfo {author} {\bibfnamefont {S.}~\bibnamefont {Sinha}},\ }\href
  {http://stacks.iop.org/0264-9381/25/i=16/a=165003} {\bibfield  {journal}
  {\bibinfo  {journal} {Class. Quant. Grav.}\ }\textbf {\bibinfo {volume}
  {25}},\ \bibinfo {pages} {165003} (\bibinfo {year} {2008})}\BibitemShut
  {NoStop}%
\bibitem [{\citenamefont {Will}\ and\ \citenamefont
  {Wiseman}(1996)}]{WillWiseman:1996}%
  \BibitemOpen
  \bibfield  {author} {\bibinfo {author} {\bibfnamefont {C.~M.}\ \bibnamefont
  {Will}}\ and\ \bibinfo {author} {\bibfnamefont {A.~G.}\ \bibnamefont
  {Wiseman}},\ }\href {\doibase 10.1103/PhysRevD.54.4813} {\bibfield  {journal}
  {\bibinfo  {journal} {Phys. Rev. D}\ }\textbf {\bibinfo {volume} {54}},\
  \bibinfo {pages} {4813} (\bibinfo {year} {1996})}\BibitemShut {NoStop}%
\bibitem [{\citenamefont {Ajith}\ \emph {et~al.}(2011)\citenamefont {Ajith},
  \citenamefont {Boyle}, \citenamefont {Brown}, \citenamefont {Fairhurst},
  \citenamefont {Hannam}, \citenamefont {Hinder}, \citenamefont {Husa},
  \citenamefont {Krishnan}, \citenamefont {Mercer}, \citenamefont {Ohme},
  \citenamefont {Ott}, \citenamefont {Read}, \citenamefont {Santamar{\'{i}}a},\
  and\ \citenamefont {Whelan}}]{AjithEtAl:2007}%
  \BibitemOpen
  \bibfield  {author} {\bibinfo {author} {\bibfnamefont {P.}~\bibnamefont
  {Ajith}}, \bibinfo {author} {\bibfnamefont {M.}~\bibnamefont {Boyle}},
  \bibinfo {author} {\bibfnamefont {D.~A.}\ \bibnamefont {Brown}}, \bibinfo
  {author} {\bibfnamefont {S.}~\bibnamefont {Fairhurst}}, \bibinfo {author}
  {\bibfnamefont {M.}~\bibnamefont {Hannam}}, \bibinfo {author} {\bibfnamefont
  {I.}~\bibnamefont {Hinder}}, \bibinfo {author} {\bibfnamefont
  {S.}~\bibnamefont {Husa}}, \bibinfo {author} {\bibfnamefont {B.}~\bibnamefont
  {Krishnan}}, \bibinfo {author} {\bibfnamefont {R.~A.}\ \bibnamefont
  {Mercer}}, \bibinfo {author} {\bibfnamefont {F.}~\bibnamefont {Ohme}},
  \bibinfo {author} {\bibfnamefont {C.~D.}\ \bibnamefont {Ott}}, \bibinfo
  {author} {\bibfnamefont {J.~S.}\ \bibnamefont {Read}}, \bibinfo {author}
  {\bibfnamefont {L.}~\bibnamefont {Santamar{\'{i}}a}}, \ and\ \bibinfo
  {author} {\bibfnamefont {J.~T.}\ \bibnamefont {Whelan}},\ }\href@noop {}
  {\enquote {\bibinfo {title} {Data formats for numerical relativity waves},}\
  } (\bibinfo {year} {2011}),\ \Eprint {http://arxiv.org/abs/0709.0093v3}
  {0709.0093v3 [gr-qc]} \BibitemShut {NoStop}%
\bibitem [{\citenamefont {Boyle}(2011)}]{Boyle:2011dy}%
  \BibitemOpen
  \bibfield  {author} {\bibinfo {author} {\bibfnamefont {M.}~\bibnamefont
  {Boyle}},\ }\href {http://link.aps.org/doi/10.1103/PhysRevD.84.064013}
  {\bibfield  {journal} {\bibinfo  {journal} {Phys. Rev. D}\ }\textbf {\bibinfo
  {volume} {84}},\ \bibinfo {pages} {064013} (\bibinfo {year}
  {2011})}\BibitemShut {NoStop}%
\bibitem [{\citenamefont {Boyle}\ \emph {et~al.}(2009)\citenamefont {Boyle},
  \citenamefont {Brown},\ and\ \citenamefont {Pekowsky}}]{Boyle2008b}%
  \BibitemOpen
  \bibfield  {author} {\bibinfo {author} {\bibfnamefont {M.}~\bibnamefont
  {Boyle}}, \bibinfo {author} {\bibfnamefont {D.~A.}\ \bibnamefont {Brown}}, \
  and\ \bibinfo {author} {\bibfnamefont {L.}~\bibnamefont {Pekowsky}},\ }\href
  {http://www.iop.org/EJ/abstract/0264-9381/26/11/114006} {\bibfield  {journal}
  {\bibinfo  {journal} {Class. Quant. Grav.}\ }\textbf {\bibinfo {volume}
  {26}},\ \bibinfo {pages} {114006} (\bibinfo {year} {2009})}\BibitemShut
  {NoStop}%
\bibitem [{\citenamefont {Shoemake}(1985)}]{Shoemake:1985}%
  \BibitemOpen
  \bibfield  {author} {\bibinfo {author} {\bibfnamefont {K.}~\bibnamefont
  {Shoemake}},\ }\href {\doibase 10.1145/325165.325242} {\bibfield  {journal}
  {\bibinfo  {journal} {{ACM} {SIGGRAPH} Computer Graphics}\ }\textbf {\bibinfo
  {volume} {19}},\ \bibinfo {pages} {245} (\bibinfo {year} {1985})}\BibitemShut
  {NoStop}%
\bibitem [{\citenamefont {Grassia}(1998)}]{Grassia:1998}%
  \BibitemOpen
  \bibfield  {author} {\bibinfo {author} {\bibfnamefont {F.~S.}\ \bibnamefont
  {Grassia}},\ }\href
  {http://citeseer.ist.psu.edu/viewdoc/summary?doi=10.1.1.132.20} {\bibfield
  {journal} {\bibinfo  {journal} {Journal of Graphics Tools}\ }\textbf
  {\bibinfo {volume} {3}},\ \bibinfo {pages} {29} (\bibinfo {year}
  {1998})}\BibitemShut {NoStop}%
\bibitem [{\citenamefont {Boyle}\ and\ \citenamefont
  {Mrou{\'{e}}}(2009)}]{Boyle-Mroue:2008}%
  \BibitemOpen
  \bibfield  {author} {\bibinfo {author} {\bibfnamefont {M.}~\bibnamefont
  {Boyle}}\ and\ \bibinfo {author} {\bibfnamefont {A.~H.}\ \bibnamefont
  {Mrou{\'{e}}}},\ }\href {\doibase 10.1103/PhysRevD.80.124045} {\bibfield
  {journal} {\bibinfo  {journal} {Phys. Rev. D}\ }\textbf {\bibinfo {volume}
  {80}},\ \bibinfo {pages} {124045} (\bibinfo {year} {2009})}\BibitemShut
  {NoStop}%
\bibitem [{\citenamefont {Goldstein}\ \emph {et~al.}(2001)\citenamefont
  {Goldstein}, \citenamefont {Poole},\ and\ \citenamefont
  {Safko}}]{GoldsteinEtAl:2001}%
  \BibitemOpen
  \bibfield  {author} {\bibinfo {author} {\bibfnamefont {H.}~\bibnamefont
  {Goldstein}}, \bibinfo {author} {\bibfnamefont {C.~P.}\ \bibnamefont
  {Poole}}, \ and\ \bibinfo {author} {\bibfnamefont {J.~L.}\ \bibnamefont
  {Safko}},\ }\href@noop {} {\emph {\bibinfo {title} {Classical Mechanics}}},\
  \bibinfo {edition} {3rd}\ ed.\ (\bibinfo  {publisher} {Addison Wesley},\
  \bibinfo {address} {Reading, MA},\ \bibinfo {year} {2001})\BibitemShut
  {NoStop}%
\bibitem [{\citenamefont {Wigner}(1959)}]{Wigner:1959}%
  \BibitemOpen
  \bibfield  {author} {\bibinfo {author} {\bibfnamefont {E.~P.}\ \bibnamefont
  {Wigner}},\ }\href@noop {} {\emph {\bibinfo {title} {Group theory and its
  application to the quantum mechanics of atomic spectra}}}\ (\bibinfo
  {publisher} {Academic Press, Inc.},\ \bibinfo {address} {New York, NY},\
  \bibinfo {year} {1959})\BibitemShut {NoStop}%
\bibitem [{\citenamefont {Newman}\ and\ \citenamefont
  {Penrose}(1966)}]{NewmanPenrose:1966}%
  \BibitemOpen
  \bibfield  {author} {\bibinfo {author} {\bibfnamefont {E.~T.}\ \bibnamefont
  {Newman}}\ and\ \bibinfo {author} {\bibfnamefont {R.}~\bibnamefont
  {Penrose}},\ }\href {\doibase 10.1063/1.1931221} {\bibfield  {journal}
  {\bibinfo  {journal} {J. Math. Phys.}\ }\textbf {\bibinfo {volume} {7}},\
  \bibinfo {pages} {863} (\bibinfo {year} {1966})}\BibitemShut {NoStop}%
\bibitem [{\citenamefont {Boyle}(2008)}]{BoyleThesis}%
  \BibitemOpen
  \bibfield  {author} {\bibinfo {author} {\bibfnamefont {M.}~\bibnamefont
  {Boyle}},\ }\emph {\bibinfo {title} {Accurate gravitational waveforms from
  binary black-hole systems}},\ \href
  {http://etd.caltech.edu/etd/available/etd-01122009-143851/} {Ph.D. thesis},\
  \bibinfo  {school} {California Institute of Technology} (\bibinfo {year}
  {2008})\BibitemShut {NoStop}%
\bibitem [{\citenamefont {Doran}\ and\ \citenamefont
  {Lasenby}(2003)}]{DoranLasenby:2003}%
  \BibitemOpen
  \bibfield  {author} {\bibinfo {author} {\bibfnamefont {C.}~\bibnamefont
  {Doran}}\ and\ \bibinfo {author} {\bibfnamefont {A.}~\bibnamefont
  {Lasenby}},\ }\href@noop {} {\emph {\bibinfo {title} {Geometric Algebra for
  Physicists}}}\ (\bibinfo  {publisher} {Cambridge University Press},\ \bibinfo
  {address} {New York, NY},\ \bibinfo {year} {2003})\BibitemShut {NoStop}%
\end{thebibliography}%

\end{document}